\documentclass[prl,twocolumn,superscriptaddress]{revtex4}
\usepackage{color}
\usepackage{amsmath,amsthm,mathrsfs,dsfont}
\usepackage{graphicx}
\usepackage{tikz}
\usepackage{xr}
\usepackage{xc}
\usepackage{float}
\usepackage{bbm}
\usepackage{epsfig} 
\usepackage{verbatim}
\usepackage{array}
\usepackage{setspace}
\usepackage{bbding}
\usepackage{amssymb}
\usepackage{pifont}
\usepackage{braket}
\usepackage{tikz}

\usetikzlibrary{quantikz}

\usepackage{newtxtext,newtxmath}

\externalcitedocument[suppl-]{../suppl/low-depth0312}
   
\usepackage[unicode=true,
 bookmarks=true,bookmarksnumbered=false,bookmarksopen=false,
 breaklinks=false,pdfborder={0 0 1},backref=false,colorlinks=true]
 {hyperref}
\hypersetup{
 linkcolor=magenta, urlcolor=blue, citecolor=blue, pdfstartview={FitH}, hyperfootnotes=false, unicode=true}
 
\usepackage{color}

\makeatletter
\@ifundefined{textcolor}{}
{%
 \definecolor{BLACK}{gray}{0}
 \definecolor{WHITE}{gray}{1}
 \definecolor{RED}{rgb}{1,0,0}
 \definecolor{GREEN}{rgb}{0,1,0}
 \definecolor{BLUE}{rgb}{0,0,1}
 \definecolor{CYAN}{cmyk}{1,0,0,0}
 \definecolor{MAGENTA}{cmyk}{0,1,0,0}
 \definecolor{YELLOW}{cmyk}{0,0,1,0}
}

\usepackage{amsfonts}\usepackage{tabularx}\usepackage{dcolumn}\usepackage{bm}\usepackage{graphicx}\usepackage{epstopdf}

\hypersetup{urlcolor=blue}

\makeatother

\newcolumntype{C}[1]{>{\centering\arraybackslash$}p{#1}<{$}}

\usepackage{algorithm}
\usepackage{algorithmic}

\begin{document}

\widetext
\title{Unbiased random circuit compiler for time-dependent Hamiltonian simulation}

\author{Xiao-Ming Zhang}
\affiliation{Center on Frontiers of Computing Studies, Peking University, Beijing 100871, China}
\affiliation{School of Computer Science, Peking University, Beijing 100871, China}

\author{Zixuan Huo}
\affiliation{Center on Frontiers of Computing Studies, Peking University, Beijing 100871, China}
\affiliation{School of Computer Science, Peking University, Beijing 100871, China}

\author{{Kecheng Liu} 
}
\affiliation{Center on Frontiers of Computing Studies, Peking University, Beijing 100871, China}
\affiliation{School of Computer Science, Peking University, Beijing 100871, China}

\author{Ying Li}
\affiliation{Graduate School of China Academy of Engineering Physics, Beijing 100193, China}

\author{Xiao Yuan}
\email{xiaoyuan@pku.edu.cn}
\affiliation{Center on Frontiers of Computing Studies, Peking University, Beijing 100871, China}
\affiliation{School of Computer Science, Peking University, Beijing 100871, China}

\begin{abstract}
Time-dependent Hamiltonian simulation (TDHS) is a critical task in quantum computing. 
Existing algorithms are generally biased with a small algorithmic error $\varepsilon$, and the gate complexity scales as $O(\text{poly}(1/\varepsilon))$ for product formula-based methods and could be improved to be polylogarithmic with complicated circuit constructions. Here, we develop an unbiased random compiler for TDHS by combining Dyson expansion, an unbiased continuous sampling method for quantum evolution, and leading order rotations, and it is free from algorithmic errors. Our method has the single- and two-qubit gate complexity $O(\Lambda^2)$ with a constant sampling overhead, where  $\Lambda$ is the time integration of the Hamiltonian strength. We perform numerical simulations for a spin model under the interaction picture and the adiabatic ground state preparation for molecular systems. In both examples, we observe notable improvements of our method over existing ones.
Our work paves the way to efficient realizations of TDHS.
\end{abstract}
\maketitle

As one of the most important tasks in quantum information processing, Hamiltonian simulation  is a promising candidate for the first practical application of quantum computing~\cite{Childs.18}.  In particular, time-dependent Hamiltonian simulation (TDHS) can be used to explore rich physics phenomena, ranging from adiabatic quantum evolution~\cite{Albash.18} to driven systems under highly-oscillated external driving fields~\cite{Changsuk.16}. Moreover, time-independent Hamiltonians can be transformed to a time-dependent one in the interaction picture,  providing significant improvements to the performance of Hamiltonian simulation~\cite{Low.18}.

TDHS can be realized based on product formula~\cite{Huyghebaert.90,Wiebe.10,Poulin.11,Wecker.15,Berry.20,An.21}. For example, 
Wiebe~\textit{et.~al.}~developed higher-order product formula algorithms for sufficiently smooth Hamiltonians~\cite{Wiebe.10}.  
Poulin~\textit{et.~al.}~developed a Monte Carlo method based on the time-average of the Hamiltonian, achieving gate count being quadratic to the operator norm of the Hamiltonian and independent of the Hamiltonian derivative~\cite{Poulin.11}. Berry~\textit{et.~al.}~developed a continuous qDRIFT (c-qDRIFT) method by taking the instantaneous norm of Hamiltonian into consideration~\cite{Berry.20}. 
Their gate count has $L^{1}$ norm scaling, and is independent of the number of terms in the Hamiltonian. 
An~\textit{et.~al.}~generalized the product formula methods to unbounded Hamiltonian, and achieved vector norm gate count scaling~\cite{An.21}. 
However, methods above under the product formula framework all have $O(\text{poly}(1/\varepsilon))$ gate count scaling for algorithmic error $\varepsilon$, restricting the precision that can be achieved when the circuit depth is limited. 
The complexity can be improved to be polylogarithmic with algorithms beyond the product formula framework~\cite{Berry.14,Low.18,Kieferov.19,Berry.20,Chen.21,An.22,Watkins.22,Mizuta.22,Rajput.22}. But these methods require ancillary qubits and oracles based on multi-qubit entangling gates, so their implementations are actually more challenging for intermediate-scale problems and near-term quantum devices, as shown in Ref.~\cite{Childs.18}.

All existing TDHS algorithms are approximate and hence biased. We have to increase the circuit depth to reduce the algorithmic error $\varepsilon$. 
Here, we propose an unbiased random circuit compiler (URCC) for general time-dependent Hamiltonian simulation. The single- and two-qubit gate count of our method is $O(\Lambda^2)$, where $\Lambda$ is the time integration of total Hamiltonian strength. In particular, the gate count is independent of the accuracy and number of terms in the Hamiltonian. Moreover, our method is compatible with simultaneous measurement techniques~\cite{Huang.22,Verteletskyi.20}. We provide two numerical examples, the spin model in the interaction picture, and adiabatic ground state preparation for molecular systems. In both examples, we observe significant gate count reduction compared to the existing method.

 \vspace{0.2cm}
 
\begin{figure}[t]
    \centering
          \includegraphics[width=1\columnwidth]{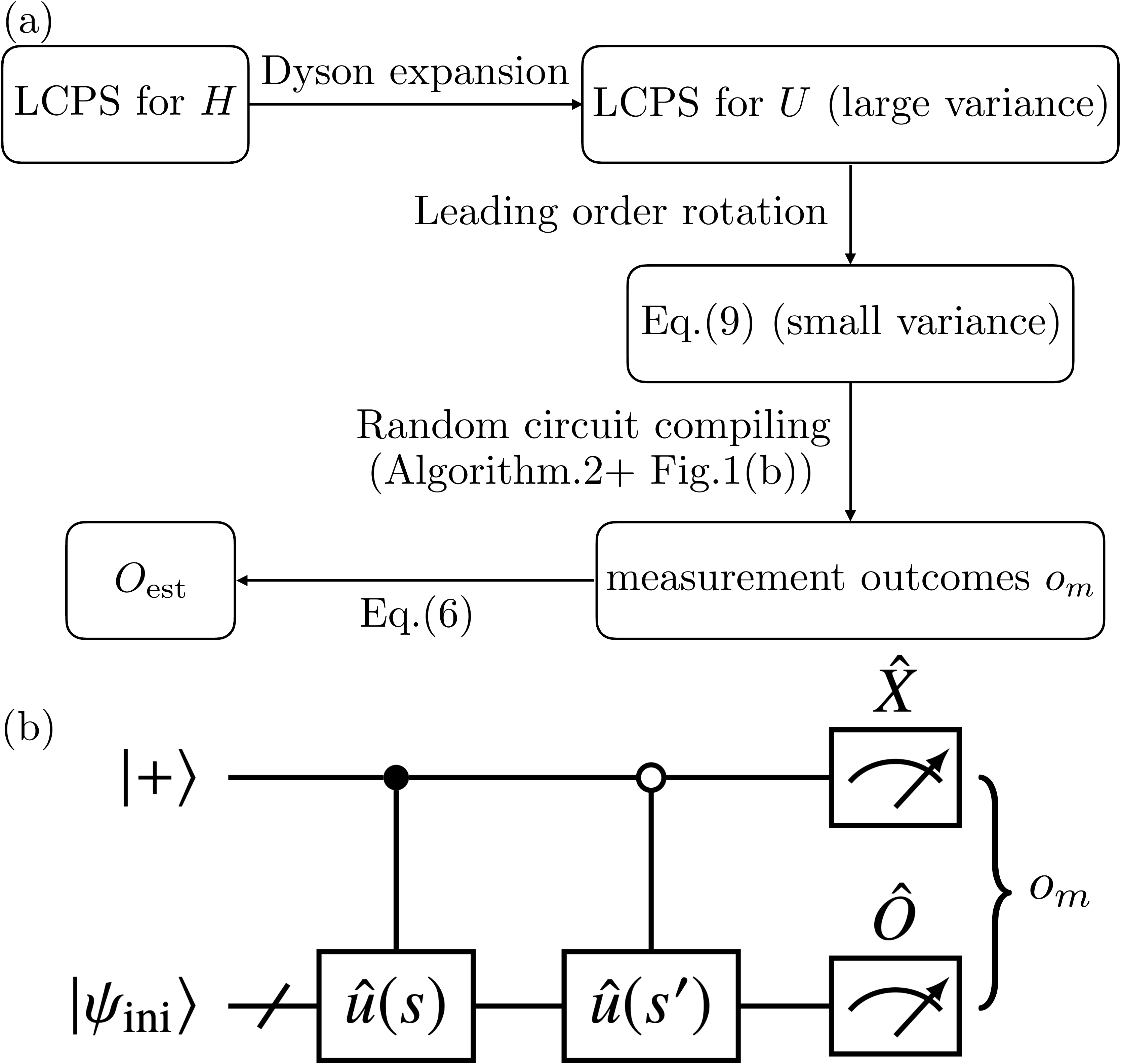}
       \caption{(a) Overview of the URCC algorithm for short-time evolution. (c) Sketch of the random quantum circuit with output $o_m$.  }
       \label{fig:1}
\end{figure}

\textbf{\textit{Overview.---}} 
Given a time-dependent Hamiltonian $H(t)$ and quantum state $|\psi_{0}\rangle$, the quantum state after time $\tau>0$ is $|\psi_\tau\rangle=U(0,\tau)|\psi_0\rangle$, where 
\begin{align}\label{eq:DY}
U(0,\tau)=\mathcal{T}\exp\left[-i\int_{0}^{\tau} dtH(t)\right].
\end{align}
Here, $\mathcal{T}$ is the time-ordering operator. Instead of $|\psi_{\tau}\rangle$ \textit{per se}, we care about the measurement outcome of observable $\hat O$, whose expectation value at time $\tau$ is

\begin{align}\label{eq:O}
\langle O\rangle=\text{tr}\left(\hat O U(0,\tau)|\psi_0\rangle\langle\psi_0|U(0,\tau)^\dag\right).
\end{align}
Our algorithm outputs an unbiased estimator for $\langle O\rangle$ by the combination of three different techniques: Dyson expansion of time-dependent evolution~\cite{Dyson.49}, classical unbiased continuous sampling of the linear combination of unitaries~\cite{Faehrmann.22}, and leading order rotation~\cite{Yang.21}. 
An overview of our algorithm is also provided in Fig.~\ref{fig:1}(a). 
Based on the Dyson expansion to infinite orders, we first rewrite the evolution as the linear combination of Pauli strings (LCPS). Then, we develop an unbiased and efficient circuit sampling algorithm according to the LCPS. The variance of such  LCPS-based sampling is, however, exponential with respect to the time integral of Hamiltonian strength. We then apply the leading order rotation technique, which combines the zero and the first order of the Dyson expansion into a rotation operator with $O(n)$ circuit depth. This reduces the variance of unbiased sampling from exponential to polynomial. 

We note that while some techniques above have been separately discussed in the literature for different purposes, our work for the time studies their combinations, which are essential for our URCC.

 \vspace{0.2cm}
 
\textbf{\textit{Dyson expansion.---}} 
We consider an $n$-qubit Hamiltonian written as the linear combination of Pauli strings 
\begin{align}\label{eq:ham}
H(t)=\sum_{p=1}^Ph_p(t)\hat \sigma_p
\end{align}
 with $h_p(t)\geqslant0$, $\hat \sigma_p\in\mathbb{P}_{\pm}\equiv\{\pm p|p\in\mathbb{P}\}$ and $\mathbb{P}\equiv\{\hat I,\hat X,\hat Y,\hat Z\}^{\otimes n}$. Here $\hat I,\hat X,\hat Y,\hat Z$ are single qubit identity and Pauli operators. We begin with a short-time evolution from $\tau_{\text{ini}}$ to $\tau_{\text{ed}}$, and discuss the generalization to long-time evolution later. We define the total Hamiltonian strength $h_{\text{tot}}(t)=\sum_{p=1}^{P}h_p(t)$, the time integral of the strength of each term $\lambda_p=\int_{\tau_{\text{ini}}}^{\tau_{\text{ed}}}h_{p}(t)dt$, and its time integration $\lambda=\sum_{p=1}^{P}\lambda_p$. 
 
We simply denote $U=U(\tau_{\text{ini}},\tau_{\text{ed}})$. According to Dyson expansion, $U$ can be expressed as the following LCU forms (see App.~\ref{app:dy})
\begin{subequations}\label{eq:DY}
\begin{align}
U&=e^{\lambda}\sum_{l=0}^{\infty}\text{Poi}(\lambda;l)D(l),\label{eq:Dya}\\
D(l)&=\sum_{p_1,\cdots,p_l=1}^{P}\text{Pdy}(l;\bm{p})\hat P(\bm{p}),\label{eq:Dyb}%
\end{align}
where $\text{Poi}(\lambda;l)=\big(\frac{\lambda^{l}}{l!}\big)/e^{\lambda}$ is a Poisson probability distribution. $D(l)$ is the rescaled $l$th-order Dyson series
with $\bm{p}=[p_1,\cdots,p_l]$ and $\hat P(\bm{p})\equiv(-i\hat \sigma_{p_1})(-i\hat \sigma_{p_2})\cdots(-i\hat \sigma_{p_l})$. Moreover, we have
\begin{align}\label{eq:prdy}
\text{Pdy}(l;\bm{p})=\frac{l!}{\lambda^l}\left(\mathcal{T}\int_{t_{\text{ini}}}^{t_{\text{ed}}}d\bm{t} \prod_{l'=1}^lh_{p_{l'}}(t_{l'})\right).
\end{align}
\end{subequations}
Here, $\mathcal{T}\int_{t_{\text{ini}}}^{t_{\text{ed}}}d\bm{t}\equiv\int_{\tau_{\text{ini}}\leqslant t_l\leqslant t_{l-1}\leqslant  \cdots\leqslant t_1\leqslant\tau_{\text{ed}}}dt_l\cdots dt_1$ is the time-ordered integration. We note that Eq.~\eqref{eq:prdy} is normalized, i.e. $\sum_{\bm{p}}\text{Pdy}(l;\bm{p})=1$ when $h_p(t)\geqslant0$, which is ensured by our definition (see App~\ref{app:dy}). So $\text{Pdy}(l;\bm{p})$ can be considered as the distribution of Pauli strings $\hat P(\bm{p})$.

Having provided a linear combination of unitary (LCU) for $U$, in below, we show how $\langle O\rangle$ can be unbiasedly estimated by random circuit sampling according to its LCU. 

We suppose a unitary can be decomposed as 
\begin{align} \label{eq:normf}
U=C\sum_{s}c(s)\hat u(s),
\end{align} 
where $c(s)\geqslant0$, $\sum_{s}c(s)=1$, and $\hat u(s)$ are unitaries. $C>0$ is the normalization factor for LCU. The circuit used in our protocol is shown in Fig.~\ref{fig:1}(b)~\cite{Faehrmann.22}. We introduce an ancillary qubit, and initialize the quantum state to $|+\rangle\langle+|\otimes\rho_{\text{ini}}$, where $\rho_{\text{ini}}=|\psi_{\text{ini}}\rangle\langle\psi_{\text{ini}}|$. We then apply the unitary $|0\rangle\langle0|\otimes\hat u(s')+|1\rangle\langle1|\otimes\hat u(s)$, and perform measurement at observable $\hat X\otimes \hat O$. The expectation value of the measurement outcome is $\frac{1}{2}\text{tr}\left( O \left(\hat u(s)\rho_{\text{ini}}\hat u(s')^\dag+\hat u(s')\rho_{\text{ini}}\hat u(s)^\dag\right)\right)$. If we sample $u(s)$ and $u(s')$ according to the probability $c(s)$ and $c(s')$ respectively, the measurement outcome $o$ satisfies $C^2\overline o=\text{tr}(U\rho_{\text{ini}} U^\dag)=\langle O\rangle$. 
We repeat the sampling for $M$ times, and denote the measurement outcome at the $m$th trial as $o_m$. The estimator for $\langle O\rangle$ is given by
\begin{align}\label{eq:oest}
O_{\text{est}}=\sum_{m=1}^{M}\frac{C^2}{M}o_m,
\end{align}
which satisfies $\overline{O_{\text{est}}}=\langle O\rangle$. Moreover, we have $o_m\in[-\|\hat O\|,\|\hat O\|]$ where $\|\cdot\|$ is the spectral norm. According to Hoeffding's bound, with failure probability $\delta$, the sampling error satisfies $|O_{\text{est}}-\langle O\rangle|\leqslant\varepsilon_{\text{samp}}$, where~\cite{Faehrmann.22,Pei.22} 
\begin{align}\label{eq:err}
\varepsilon_{\text{samp}}=\big\|\hat O\big\|C^2\sqrt{2\ln(2/\delta)/M}.
\end{align}

The random circuit sampling protocol above is general. For Eq.~\eqref{eq:DY}, the corresponding normalization factor $C$, probability $c(s)$ and elementary unitaires $\hat u(s)$ are $e^\lambda$, $\text{Poi}(\lambda;l)\text{Pdy}(l;\bm{}p)$ and $\hat P(\bm{p})$ respectively. Note that $\hat P(\bm{p})$ is a single layer of single-qubit gates that can easily be implemented. Therefore, the remaining task is to perform a random sampling of $\bm{p}$ according to $\text{Poi}(\lambda;l)\text{Pdy}(l;\bm{p})$.

 The sampling can be separated into two stages. In stage $1$, we sample the order $l$ according to Poisson distribution $\text{Poi}(\lambda;l)$. In stage $2$, we sample the Pauli string $\hat P(\bm{p})$ according to $\text{Pdy}(l;\bm{p})$. Stage $1$ is simple, while the main challenge lies in stage $2$.
A naive protocol for stage $2$ is to calculate the integration in Eq.~\eqref{eq:prdy} to obtain a discrete distribution of $\hat P(\bm{p})$. But this protocol requires heavy classical calculations with runtime increasing exponentially with $l$. Instead, we transfer stage $2$ to a continuous sampling problem as follows, which significantly reduces the sampling time.

 \vspace{0.2cm}
 
\textbf{\textit{Unbiased continuous sampling.---}} 
To begin with, we notice that $\text{Pdy}(l;\bm{p})\hat P(\bm{p})$ can be rewritten as
\begin{subequations}\label{eq:prod2}
\begin{align}
\text{Pdy}(l;\bm{p})\hat P(\bm{p})&=\frac{l!}{\lambda^l}\left( \mathcal{T}\int_{t_{\text{ini}}}^{t_{\text{ed}}}d\bm{t} \prod_{l'=1}^lh_{\text{tot}}(t_{l'}) Q(t_{l'})\right),\label{eq:prod2a}\\
 Q(t_{l'})&=\sum_{p=1}^{P}\frac{h_{p}(t_{l'})}{h_{\text{tot}}(t_{l'})}(-i\hat\sigma_p).\label{eq:prod2b}
\end{align}
\end{subequations}
Eq.~\eqref{eq:prod2a}, can be considered as the linear combination of $Q(t_0)Q(t_1)\cdots Q(t_l)$ of all $\bm{t}$ satisfying $\tau_{\text{ini}}\leqslant t_l\leqslant t_{l-1}\leqslant  \cdots\leqslant t_1\leqslant\tau_{\text{ed}}$, weighted by the coefficient proportional to $\prod_{l'=1}^lh_{\text{tot}}(t_{l'})$. Moreover, each $Q(t_{l'})$ can be considered as the linear combination of $-i\hat\sigma_p$ weighted by the coefficient $\frac{h_{p}(t_{l'})}{h_{\text{tot}}(t_{l'})}$ (Eq.~\eqref{eq:prod2b}). Because $\hat P(\bm{p})=(-i\hat \sigma_{p_1})(-i\hat \sigma_{p_2})\cdots(-i\hat \sigma_{p_l})$, stage 2 can be realized by the following two substeps (see also Alg.~\ref{alg:pdy}).

In the first substep, for each $l'=1,2,\cdots,l$, we sample time $\tilde t_{l'}\in[t_{\text{ini}}, t_{\text{ed}}]$ independently with probability $\text{Pr}[\tilde t_{l'}=t]\propto h_{\text{tot}}(t)$. Then, we re-order the sampled moments from large to small and obtain the vector $\bm{t}=[t_1,\cdots,t_l]$, such that $t_1\geqslant\cdots\geqslant t_{l}$. In this way, we obtain time vector $\bm{t}$ of descending order with probability proportional to $\prod_{l'=1}^{l}h_{\text{tot}}(t_{l'})$. In this way, we have realized the sampling of $Q(t_0)Q(t_1)\cdots Q(t_l)$. In the second substep, for each $ l'\in\{1,\cdots, l\}$, we sample $p_{l'}$ according to $\text{Pr}(p_{l'})=h_p(t_{l'})/h_{\text{tot}}(t_{l'})$, and finally output the operator $\hat P(\bm{p})$.

We have shown how Pauli strings can be efficiently sampled for Eq.~\eqref{eq:DY}. We also note that $\hat P(\bm{p})$, after simplification, is a layer of single-qubit gates that can be simulated in classical computers efficiently. So it also represents a classical algorithm of independent interest for small $\lambda$. However, the normalization factor for Eq.~\eqref{eq:DY} is $e^\lambda$. According to Eq.~\eqref{eq:err}, the error increases exponentially with $\lambda$, which is inefficient for long-time or strong Hamiltonian strength evolution.

Below, we show how to reduce the sampling error with a more sophisticated LCU, which is not classically efficiently stimulable in general.

\begin{algorithm} [t]
\caption{Sample $\hat P(\bm{p})$ according to Pdy$(l;\bm{p})$\label{alg:pdy}}  
\begin{algorithmic}[t]

\STATE\textbf{for} $l'=1$ to $l$:
\STATE \quad Sample $\tilde t_{l'}\in[t_{\text{ini}},t_{\text{ed}}]$ with $\text{Pr}[\tilde t_{l'}=t]\propto h_{\text{tot}}(t)$
\STATE\textbf{end for}
\STATE re-order $\{\tilde t_{l'}\}$ as $t_{l}<t_{l-1}<\cdots <t_1$

\STATE\textbf{for} $l'=1$ to $l$:
\STATE\quad Sample $p_{l'}$ with  $\text{Pr}[p_{l'}=p]\propto h_{p}(t_{l'})$
\STATE\textbf{end for}

\STATE \textbf{Output} $(-i)^{l}\hat\sigma_{p_1}\hat\sigma_{p_2}\cdots\hat\sigma_{p_{l}}$
\end{algorithmic} 
\end{algorithm}

\begin{algorithm} [t]
\caption{Unitary sampling according to Eq.~\eqref{eq:ULR}\label{alg:2}}  
\begin{algorithmic}[t]

\STATE Set $A=L$ (or $A=R$) with probability $\frac{C_{L}}{C_L+C_R}$ $\left(\frac{C_{R}}{C_L+C_R}\right)$
\STATE \textbf{if} $A=L$:
\STATE \quad Sample $p$ with $\text{Pr}[p]=\lambda_p$
\STATE \quad \textbf{Output} $\exp(-i\phi\hat\sigma_p)$ 
\STATE \textbf{else}:
\STATE \quad Sample $l$ with $\text{Pr}[l]=\text{Poi}(\lambda;l)$
\STATE \quad \textbf{while} $l<2$:
\STATE \quad \quad Sample $l$ with $\text{Pr}[l]=\text{Poi}(\lambda;l)$
\STATE \quad \textbf{end while} 
\STATE \quad Sample $\hat P(\bm{p})$ according to Pdy$(l;\bm{p})$
\STATE \quad \textbf{Output} $\hat P(\bm{p})$ 
\STATE \textbf{end if} 

\end{algorithmic} 
\end{algorithm}

 \vspace{0.2cm}
 
\textbf{\textit{Leading order rotation.---}} 
The expression of $U$ in Eq.~\eqref{eq:Dya} can be divided into leading order and remaining order terms separated by $l<2$ and $l\geqslant2$:
\begin{subequations}\label{eq:ULR}
\begin{align}
U=L+R,
\end{align}
where the remaining order term is
\begin{align}\label{eq:ulrr}
R=e^{\lambda}\sum_{l=2}^{\infty}\sum_{p_1,\cdots,p_l=1}^{P}\text{Poi}(\lambda;l)\text{Pdy}(l;\bm{p})\hat P(\bm{p}).
\end{align}
The leading order term, after some simplifications, can be rewritten as $L=\hat{\mathbb{I}}-i\sum_{p=1}^{P}\lambda_p \hat \sigma_p$, where $\hat{\mathbb{I}}\equiv \hat I^{\otimes n}$.
Because $\hat\sigma_{p}^2=\hat{\mathbb{I}}$,  we can apply the Euler's formula to $L$. The leading order can then be rewritten as 
\begin{align}\label{eq:ulr}
L=\sum_{p=1}^P\alpha_p\exp\left(-i\phi\hat \sigma_p\right),
\end{align}
\end{subequations}
where $\phi=\arctan\left(\lambda\right)$ and  $\alpha_p=\lambda_p/\sin\phi$. If we rewrite Eq.~\eqref{eq:ULR} to a normalized form in Eq.~\eqref{eq:normf}, the normalization factor is reduced to (see App.~\ref{app:ULR})
\begin{align}\label{eq:clor}
C_{\text{lor}}=\sqrt{1+\lambda^2}+e^\lambda-1-\lambda=1+O(\lambda^2),
\end{align}
which increases only quadratically with $\lambda$.

The random circuit sampling process according to Eq.~\eqref{eq:ULR} is illustrated in Alg.~\ref{alg:2}. In the first step, we  choose $L$ or $R$ with probability $C_L/(C_L+C_R)$ or $C_L/(C_L+C_R)$. If $L$ is chosen, we sample the rotation $\exp\left(-i\phi\hat \sigma_p\right)$ with probability proportional to $\lambda_p$ (because $\alpha_p\propto\lambda_p$). If $R$ is chosen, we sample $\hat P(\bm{p})$ according to Eq.~\eqref{eq:ulrr} with two steps. In step the first step, we sample $l$ with probability proportional to Poi$(\lambda;l)$, but excluding $l= 0,1$. To do so, we perform sampling according to Poisson distribution Poi$(\lambda;l)$, and accept the result only if $l\geqslant2$. Otherwise, the result is rejected. This process is repeated until the result is accepted. In the second step, we sample $\hat P(\bm{p})$ according to Pdy$(l;\bm{p})$ with Alg.~\ref{alg:pdy}.

 \vspace{0.2cm}
 
\textbf{\textit{Long-time evolution.---}} 
For long-time evolution, we can divide the total evolution into $N_{\text{seg}}$ segment, and perform the decomposition for each segments. In this way, the normalization factor can be controlled to a constant level with a sufficiently large $N_{\text{seg}}$. For evolution from $t=0$ to $t=\tau$, we may divide it into $N_{\text{seg}}$ segments, i.e. 
$U(0,\tau)=U(0,\tau_1)U(\tau_1,\tau_2)\cdots U(\tau_{N_{\text{seg}}-1},\tau_{N_{\text{seg}}})$, where $0<\tau_1<\cdots<\tau_{N_{\text{seg}}-1}<\tau_{N_{\text{seg}}}=\tau$. The length of each segment is set such that $\int_{\tau_{j-1}}^{\tau_{j}} h_{\text{tot}}(t) dt=\lambda$ is a small constant. We then decompose the unitary at each segment in the form of Eq.~\eqref{eq:ULR}, with the normalization factor $C_{\text{lor}}$ defined in Eq.~\eqref{eq:clor}. The random circuit sampling for long-time evolution $U(0,\tau)$ can be realized by performing sampling for each segment independently, and then concatenating the output for each segment sequentially. In this way, the total normalization factor for long-time evolution is $C={C_{\text{lor}}}^{N_{\text{seg}}}=1+O(N_{\text{seg}}\lambda^2)$. We define the long-time integration of the total Hamiltonian strength as $\Lambda=\int_0^\tau h_{\text{tot}}(t) dt$, which satisfies $\Lambda=N_{\text{seg}}\lambda$. Combining with Eq.~\eqref{eq:err} and Eq.~\eqref{eq:clor}, the error satisfies
\begin{align}\label{eq:err_NM}
\varepsilon_{\text{samp}}=O\left(\frac{1}{\sqrt{M}}
\left(1+O(\Lambda^2/N^2_{\text{seg}})\right)^{2N_{\text{seg}}}\right).
\end{align}
More details about long-time evolution are provided in App.~\ref{app:long}.

According to Eq.~\eqref{eq:err_NM}, to control the sampling error to below $\varepsilon_{\text{samp}}$, it suffices to set $M=O(1/\varepsilon_{\text{samp}}^2)$ and $N_{\text{seg}}=O(\Lambda^2)$.
Because each segment $U(\tau_{j},\tau_{j+1})$ can be realized by $O(n)$ single- and two-qubit gates, the gate count of our algorithm is $N_{\text{gate}}=O(\Lambda^2n)$. If we further assume that the Hamiltonian is $k$-local ($\hat \sigma_p$ is the tensor product of $\hat I$ and no more than $k$ Pauli operators), the gate count becomes 
\begin{align}
N_{\text{gate}}=O(\Lambda^2k).
\end{align} 
We note that the gate count above assumes ideal single- and two-qubit gates. If we consider the setting of error-corrected computation based on Clifford+$T$ gates, there is an extra algorithmic error $\varepsilon_{\text{alg}}$ due to the Clifford+$T$ decomposition, and the gate count becomes $O(\Lambda^2(k+\text{log}(\Lambda /\varepsilon_{\text{alg}})))$ (see App.~\ref{app:ct}).

\begin{figure}[t]
    \centering
          \includegraphics[width=1\columnwidth]{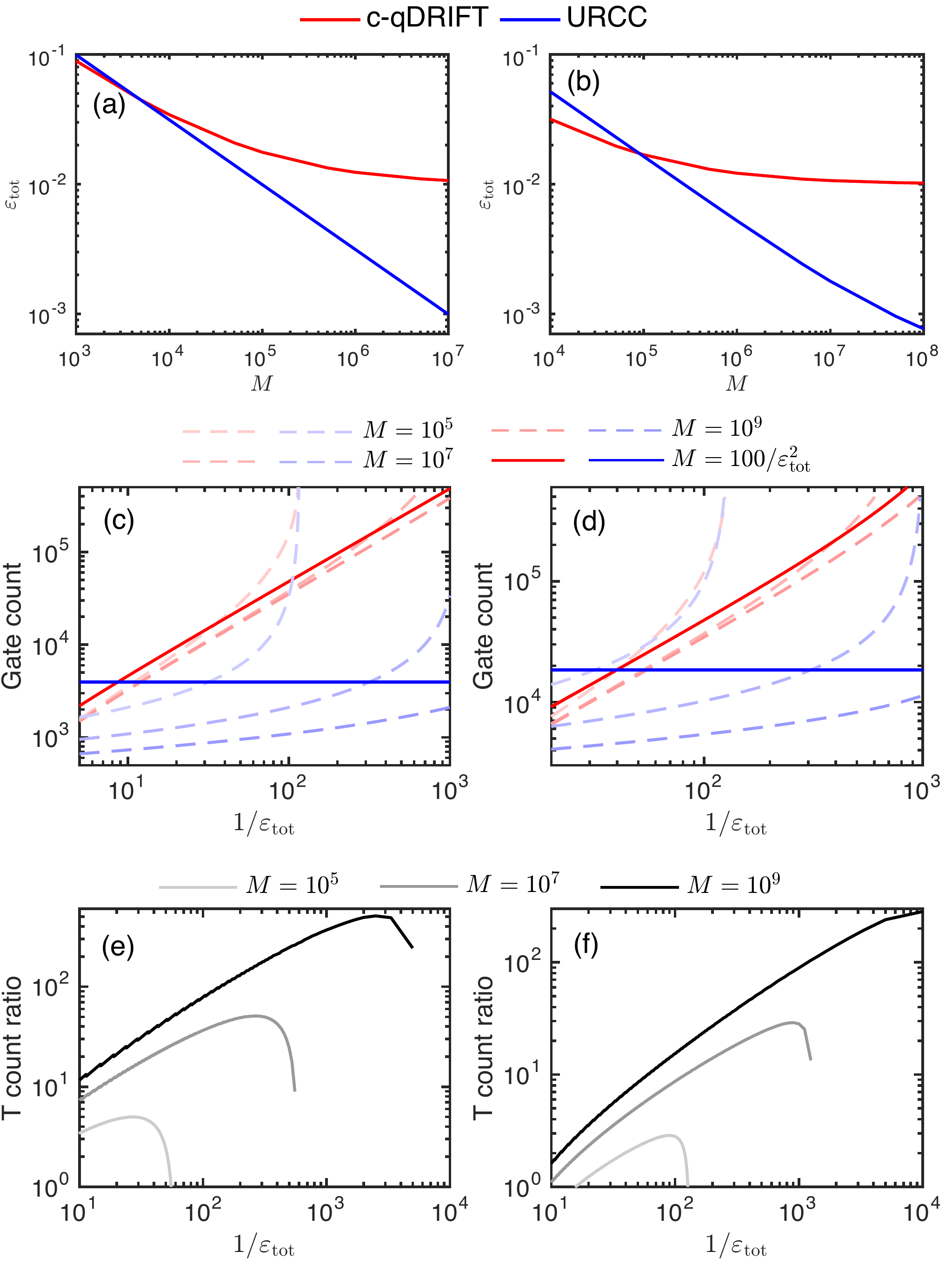}
       \caption{Numerical simulation results. (a) and (b): Error versus the number of measurements. (c) and (d): Single- and two-qubit gate counts for different error levels. Red lines represent the c-qDRIFT, and blue lines the URCC method. (e) and (f): The $T$ count ratios for c-qDRIFT method to URCC method. (a), (c) and (e) correspond to the spin model in interaction picture corresponding to Eq.~\eqref{eq:spin}. (b), (d) and (f) correspond to the adiabatic ground state preparation for $H_2$.}
       \label{fig:num}
\end{figure} 

\vspace{0.2cm}
 \textbf{\textit{Examples.---}} 
We first consider a many-body spin model under interaction picture.   With $\hat X_k,\hat Y_k,\hat Z_k$ the Pauli operators for the $k$th spin, the Hamiltonian we study is given by~\cite{Watkins.22} $H=H_0+H_{\text{int}}$, where $H_0=\omega/2\sum_{k=1}^{n}(-1)^k\hat Z_k$ and $H_{\text{int}}=J/2\sum_{k=1}^{n-1}\hat X_{k}\hat X_{k+1}+\hat Y_{k}\hat Y_{k+1}$. We transfer to the interaction picture with $\tilde{H}(t)=e^{iH_0t}H_{\text{int}}e^{-iH_0t}$, and obtain the time-dependent Hamiltonian 
\begin{align}\label{eq:spin}
\tilde H(t)=J/2\left(\cos(2\omega t)G_1+\sin(2\omega t)G_2\right),
\end{align}
where $G_1=\sum_{k=1}^{n-1}\hat X_k\hat X_{k+1}+\hat Y_k\hat Y_{k+1}$ and $G_2=\sum_{k=1}^{n-1}(-1)^{k}(\hat X_k\hat Y_{k+1}-\hat Y_k\hat X_{k+1})$. In the interaction picture, $\Lambda$ is reduced  significantly when $J\ll \omega$.   

In our simulation, we set $n=3$, initial state $|101\rangle$, observable $\hat O=\hat X_1$, and evolution time $\tau=\pi$. We estimate the sampling error according to Eq.~\eqref{eq:err} with failure probability $\delta=0.05$. We compare our unbiased random circuit compiling algorithm to the c-qDRIFT method~\cite{Berry.20}, which randomly sample $-i\lambda\sigma_p$ for evolution with probability distribution proportional to $h_p(t)$ (see~App.~\ref{app:qd}). 

Fig.~\ref{fig:num}(a) shows the error versus $M$ under fixed single- and two-qubit gate count for both algorithms.  The error of our method always reduces with increasing $M$. On the other hand, the error for c-qDRIFT converges to a fixed value. In particular, we observe $>10$ times reduction of the error with $M>10^7$.

In Fig.~\ref{fig:num}(c), we further demonstrate the gate counts required for achieving given error levels. We denote $\varepsilon_{\text{tot}}$ as total error. For URCC method, we have $\varepsilon_{\text{tot}}=\varepsilon_{\text{samp}}$; for c-qDRIFT, $\varepsilon_{\text{tot}}$ contains both sampling and algorithmic errors. More details are provided in App.~\ref{app:num}. With $M=1/\varepsilon_{\text{tot}}^2$, the gate count of our URCC method is a constant, while the gate count for c-qDRIFT increases linearly with $\varepsilon_{\text{tot}}$. Moreover, even with fixed $M$ (dashed lines), significant gate count reduction can still be observed. We note that when $M$ is fixed, $1/\varepsilon_{\text{tot}}$ converge to specific values for both methods even if we increase the gate counts. This is because there is still a sampling error due to finite $M$.

To have a more clear picture about the gate count improvement in the fault-torelant setting, in Fig.~\ref{fig:num}(e), we demonstrate the $T$ count ratio for c-qDRIFT to URCC method under fixed $M$, where we have assumed that the error due to Clifford+$T$ decomposition is sufficiently small. Details about the Clifford+$T$ decomposition are provided in App.~\ref{app:ct}. As shown in Fig.~\ref{fig:num}(e), the $T$ count ratio is larger for larger $M$. Moreover, with fixed $M$, there is a maximum of $1/\varepsilon_{\text{tot}}$ for both methods, and the ratio first increases with $1/\varepsilon_{\text{tot}}$ and then drops when getting close to the maximum of $1/\varepsilon_{\text{tot}}$.

The gate count of URCC method is independent of the number of terms in Hamiltonian. So it is suitable for models with large $P$, such as the molecular systems. Here, we take the adiabatic ground state preparation for $H_2$~\cite{Aspuru.05} as an example, although our method has much broader applications for other problems, such as chemical reactions~\cite{McArdle.20,Berry.20}. We use Jordan-Wigner transformation for $H_2$ molecule with the basis set STO-3G (App.~\ref{app:num}).  

The observable for energy measurement is the Hamiltonian itself, which should be decomposed into Pauli strings. When two Pauli strings are compatible with each other, they can be measured simultaneously. Advanced measurement techniques, such as classical shadow~\cite{Huang.22} and grouping measurements~\cite{Verteletskyi.20,Wu.21}, has been based on this property. The URCC is compatible with these methods. In this work, we use the grouping measurement for our energy measurements. We also provide a tighter bound (compared to Eq.~\eqref{eq:err}) for sampling errors in App.~\ref{app:group}. 

In Fig.~\ref{fig:num}(b), we demonstrate the total error versus $M$ for both methods. Our method also shows significant improvement. Different from the spin model, both method contains a systematic error due to the non-adiabaticity introduced by finite evolution time ($3.0\times10^{-4}$), so URCC will also converge for sufficiently large $M$. In Fig.~\ref{fig:num}(d) we demonstrate the relation between gate count and total errors. Similar to the spin model, the gate count for URCC is independent on $\varepsilon_{\text{tot}}$, while the gate count for c-qDRIFT is increases linearly with $1/\varepsilon_{\text{tot}}$. The $T$ count ratio under fixed $M$ is illustrated in Fig.~\ref{fig:num}(f), which also shows a similar trend to the spin model example.

\vspace{0.2cm}
 \textbf{\textit{Discussions.---}} We have developed an unbiased Hamiltonian simulation algorithm for TDHS. Our algorithm is general, and compatible with simultaneous measurement techniques.

Our method is suitable for scenarios when increasing the repetition times measurement is relatively simple, while the circuit depth is limited. In this case, the accuracy can be improved by simply increasing the sampling size. Moreover, because the gate count of our method is independent of the number of terms $P$ in the Hamiltonian, it is suitable for systems when $P$ are large, while most of the terms have small contribution to the total Hamiltonian strength. A typical example is molecular systems.
 
We note that the $\Lambda$ dependency of our method is not yet optimal. The combination of our method with product formula can potentially improve the $\Lambda$ scaling, either based on Magnus expansion~\cite{Magnus.54} or direct Trotter-Suzuki decomposition for smooth Hamiltonian~\cite{Wiebe.10}. But one of the challenges is that the classical compiling of quantum circuits may be more difficult when the order increase, and we left the relevant discussions for future study.
 
\textbf{\textit{Acknowledgement.---}}  This work is supported by the National Natural Science Foundation of China (Grant No. 12175003, No. 11875050, and No. 12088101), NSAF (Grant No. U1930403), and Emerging Engineering Interdisciplinary-Young Scholars Project, Peking University, the Fundamental Research Funds for the Central Universities.


\begin{thebibliography}{29}%
\makeatletter
\providecommand \@ifxundefined [1]{%
 \@ifx{#1\undefined}
}%
\providecommand \@ifnum [1]{%
 \ifnum #1\expandafter \@firstoftwo
 \else \expandafter \@secondoftwo
 \fi
}%
\providecommand \@ifx [1]{%
 \ifx #1\expandafter \@firstoftwo
 \else \expandafter \@secondoftwo
 \fi
}%
\providecommand \natexlab [1]{#1}%
\providecommand \enquote  [1]{``#1''}%
\providecommand \bibnamefont  [1]{#1}%
\providecommand \bibfnamefont [1]{#1}%
\providecommand \citenamefont [1]{#1}%
\providecommand \href@noop [0]{\@secondoftwo}%
\providecommand \href [0]{\begingroup \@sanitize@url \@href}%
\providecommand \@href[1]{\@@startlink{#1}\@@href}%
\providecommand \@@href[1]{\endgroup#1\@@endlink}%
\providecommand \@sanitize@url [0]{\catcode `\\12\catcode `\$12\catcode
  `\&12\catcode `\#12\catcode `\^12\catcode `\_12\catcode `\%12\relax}%
\providecommand \@@startlink[1]{}%
\providecommand \@@endlink[0]{}%
\providecommand \url  [0]{\begingroup\@sanitize@url \@url }%
\providecommand \@url [1]{\endgroup\@href {#1}{\urlprefix }}%
\providecommand \urlprefix  [0]{URL }%
\providecommand \Eprint [0]{\href }%
\providecommand \doibase [0]{http://dx.doi.org/}%
\providecommand \selectlanguage [0]{\@gobble}%
\providecommand \bibinfo  [0]{\@secondoftwo}%
\providecommand \bibfield  [0]{\@secondoftwo}%
\providecommand \translation [1]{[#1]}%
\providecommand \BibitemOpen [0]{}%
\providecommand \bibitemStop [0]{}%
\providecommand \bibitemNoStop [0]{.\EOS\space}%
\providecommand \EOS [0]{\spacefactor3000\relax}%
\providecommand \BibitemShut  [1]{\csname bibitem#1\endcsname}%
\let\auto@bib@innerbib\@empty
\bibitem [{\citenamefont {Childs}\ \emph {et~al.}(2018)\citenamefont {Childs},
  \citenamefont {Maslov}, \citenamefont {Nam}, \citenamefont {Ross},\ and\
  \citenamefont {Su}}]{Childs.18}%
  \BibitemOpen
  \bibfield  {author} {\bibinfo {author} {\bibfnamefont {A.~M.}\ \bibnamefont
  {Childs}}, \bibinfo {author} {\bibfnamefont {D.}~\bibnamefont {Maslov}},
  \bibinfo {author} {\bibfnamefont {Y.}~\bibnamefont {Nam}}, \bibinfo {author}
  {\bibfnamefont {N.~J.}\ \bibnamefont {Ross}}, \ and\ \bibinfo {author}
  {\bibfnamefont {Y.}~\bibnamefont {Su}},\ }\bibfield  {title} {\bibinfo
  {title} {Toward the first quantum simulation with quantum speedup},\
  }\href@noop {} {\bibfield  {journal} {\bibinfo  {journal} {Proc. Natl. Acad.
  Sci.}\ }\textbf {\bibinfo {volume} {115}},\ \bibinfo {pages} {9456} (\bibinfo
  {year} {2018})}\BibitemShut {NoStop}%
\bibitem [{\citenamefont {Albash}\ and\ \citenamefont
  {Lidar}(2018)}]{Albash.18}%
  \BibitemOpen
  \bibfield  {author} {\bibinfo {author} {\bibfnamefont {T.}~\bibnamefont
  {Albash}}\ and\ \bibinfo {author} {\bibfnamefont {D.~A.}\ \bibnamefont
  {Lidar}},\ }\bibfield  {title} {\bibinfo {title} {Adiabatic quantum
  computation},\ }\href@noop {} {\bibfield  {journal} {\bibinfo  {journal}
  {Reviews of Modern Physics}\ }\textbf {\bibinfo {volume} {90}},\ \bibinfo
  {pages} {015002} (\bibinfo {year} {2018})}\BibitemShut {NoStop}%
\bibitem [{\citenamefont {Noh}\ and\ \citenamefont
  {Angelakis}(2016)}]{Changsuk.16}%
  \BibitemOpen
  \bibfield  {author} {\bibinfo {author} {\bibfnamefont {C.}~\bibnamefont
  {Noh}}\ and\ \bibinfo {author} {\bibfnamefont {D.~G.}\ \bibnamefont
  {Angelakis}},\ }\bibfield  {title} {\bibinfo {title} {Quantum simulations and
  many-body physics with light},\ }\href@noop {} {\bibfield  {journal}
  {\bibinfo  {journal} {Reports on Progress in Physics}\ }\textbf {\bibinfo
  {volume} {80}},\ \bibinfo {pages} {016401} (\bibinfo {year}
  {2016})}\BibitemShut {NoStop}%
\bibitem [{\citenamefont {Low}\ and\ \citenamefont {Wiebe}(2018)}]{Low.18}%
  \BibitemOpen
  \bibfield  {author} {\bibinfo {author} {\bibfnamefont {G.~H.}\ \bibnamefont
  {Low}}\ and\ \bibinfo {author} {\bibfnamefont {N.}~\bibnamefont {Wiebe}},\
  }\bibfield  {title} {\bibinfo {title} {Hamiltonian simulation in the
  interaction picture},\ }\href@noop {} {\bibfield  {journal} {\bibinfo
  {journal} {arXiv:1805.00675}\ } (\bibinfo {year} {2018})}\BibitemShut
  {NoStop}%
\bibitem [{\citenamefont {Huyghebaert}\ and\ \citenamefont
  {De~Raedt}(1990)}]{Huyghebaert.90}%
  \BibitemOpen
  \bibfield  {author} {\bibinfo {author} {\bibfnamefont {J.}~\bibnamefont
  {Huyghebaert}}\ and\ \bibinfo {author} {\bibfnamefont {H.}~\bibnamefont
  {De~Raedt}},\ }\bibfield  {title} {\bibinfo {title} {Product formula methods
  for time-dependent schrodinger problems},\ }\href@noop {} {\bibfield
  {journal} {\bibinfo  {journal} {Journal of Physics A: Mathematical and
  General}\ }\textbf {\bibinfo {volume} {23}},\ \bibinfo {pages} {5777}
  (\bibinfo {year} {1990})}\BibitemShut {NoStop}%
\bibitem [{\citenamefont {Wiebe}\ \emph {et~al.}(2010)\citenamefont {Wiebe},
  \citenamefont {Berry}, \citenamefont {H{\o}yer},\ and\ \citenamefont
  {Sanders}}]{Wiebe.10}%
  \BibitemOpen
  \bibfield  {author} {\bibinfo {author} {\bibfnamefont {N.}~\bibnamefont
  {Wiebe}}, \bibinfo {author} {\bibfnamefont {D.}~\bibnamefont {Berry}},
  \bibinfo {author} {\bibfnamefont {P.}~\bibnamefont {H{\o}yer}}, \ and\
  \bibinfo {author} {\bibfnamefont {B.~C.}\ \bibnamefont {Sanders}},\
  }\bibfield  {title} {\bibinfo {title} {Higher order decompositions of ordered
  operator exponentials},\ }\href@noop {} {\bibfield  {journal} {\bibinfo
  {journal} {Journal of Physics A: Mathematical and Theoretical}\ }\textbf
  {\bibinfo {volume} {43}},\ \bibinfo {pages} {065203} (\bibinfo {year}
  {2010})}\BibitemShut {NoStop}%
\bibitem [{\citenamefont {Poulin}\ \emph {et~al.}(2011)\citenamefont {Poulin},
  \citenamefont {Qarry}, \citenamefont {Somma},\ and\ \citenamefont
  {Verstraete}}]{Poulin.11}%
  \BibitemOpen
  \bibfield  {author} {\bibinfo {author} {\bibfnamefont {D.}~\bibnamefont
  {Poulin}}, \bibinfo {author} {\bibfnamefont {A.}~\bibnamefont {Qarry}},
  \bibinfo {author} {\bibfnamefont {R.}~\bibnamefont {Somma}}, \ and\ \bibinfo
  {author} {\bibfnamefont {F.}~\bibnamefont {Verstraete}},\ }\bibfield  {title}
  {\bibinfo {title} {Quantum simulation of time-dependent hamiltonians and the
  convenient illusion of hilbert space},\ }\href@noop {} {\bibfield  {journal}
  {\bibinfo  {journal} {Phys. Rev. Lett.}\ }\textbf {\bibinfo {volume} {106}},\
  \bibinfo {pages} {170501} (\bibinfo {year} {2011})}\BibitemShut {NoStop}%
\bibitem [{\citenamefont {Wecker}\ \emph {et~al.}(2015)\citenamefont {Wecker},
  \citenamefont {Hastings}, \citenamefont {Wiebe}, \citenamefont {Clark},
  \citenamefont {Nayak},\ and\ \citenamefont {Troyer}}]{Wecker.15}%
  \BibitemOpen
  \bibfield  {author} {\bibinfo {author} {\bibfnamefont {D.}~\bibnamefont
  {Wecker}}, \bibinfo {author} {\bibfnamefont {M.~B.}\ \bibnamefont
  {Hastings}}, \bibinfo {author} {\bibfnamefont {N.}~\bibnamefont {Wiebe}},
  \bibinfo {author} {\bibfnamefont {B.~K.}\ \bibnamefont {Clark}}, \bibinfo
  {author} {\bibfnamefont {C.}~\bibnamefont {Nayak}}, \ and\ \bibinfo {author}
  {\bibfnamefont {M.}~\bibnamefont {Troyer}},\ }\bibfield  {title} {\bibinfo
  {title} {Solving strongly correlated electron models on a quantum computer},\
  }\href@noop {} {\bibfield  {journal} {\bibinfo  {journal} {Phys. Rev. A}\
  }\textbf {\bibinfo {volume} {92}},\ \bibinfo {pages} {062318} (\bibinfo
  {year} {2015})}\BibitemShut {NoStop}%
\bibitem [{\citenamefont {Berry}\ \emph {et~al.}(2020)\citenamefont {Berry},
  \citenamefont {Childs}, \citenamefont {Su}, \citenamefont {Wang},\ and\
  \citenamefont {Wiebe}}]{Berry.20}%
  \BibitemOpen
  \bibfield  {author} {\bibinfo {author} {\bibfnamefont {D.~W.}\ \bibnamefont
  {Berry}}, \bibinfo {author} {\bibfnamefont {A.~M.}\ \bibnamefont {Childs}},
  \bibinfo {author} {\bibfnamefont {Y.}~\bibnamefont {Su}}, \bibinfo {author}
  {\bibfnamefont {X.}~\bibnamefont {Wang}}, \ and\ \bibinfo {author}
  {\bibfnamefont {N.}~\bibnamefont {Wiebe}},\ }\bibfield  {title} {\bibinfo
  {title} {Time-dependent hamiltonian simulation with $l^1$-norm scaling},\
  }\href@noop {} {\bibfield  {journal} {\bibinfo  {journal} {Quantum}\ }\textbf
  {\bibinfo {volume} {4}},\ \bibinfo {pages} {254} (\bibinfo {year}
  {2020})}\BibitemShut {NoStop}%
\bibitem [{\citenamefont {An}\ \emph {et~al.}(2021)\citenamefont {An},
  \citenamefont {Fang},\ and\ \citenamefont {Lin}}]{An.21}%
  \BibitemOpen
  \bibfield  {author} {\bibinfo {author} {\bibfnamefont {D.}~\bibnamefont
  {An}}, \bibinfo {author} {\bibfnamefont {D.}~\bibnamefont {Fang}}, \ and\
  \bibinfo {author} {\bibfnamefont {L.}~\bibnamefont {Lin}},\ }\bibfield
  {title} {\bibinfo {title} {Time-dependent unbounded hamiltonian simulation
  with vector norm scaling},\ }\href@noop {} {\bibfield  {journal} {\bibinfo
  {journal} {Quantum}\ }\textbf {\bibinfo {volume} {5}},\ \bibinfo {pages}
  {459} (\bibinfo {year} {2021})}\BibitemShut {NoStop}%
\bibitem [{\citenamefont {Berry}\ \emph {et~al.}(2014)\citenamefont {Berry},
  \citenamefont {Childs}, \citenamefont {Cleve}, \citenamefont {Kothari},\ and\
  \citenamefont {Somma}}]{Berry.14}%
  \BibitemOpen
  \bibfield  {author} {\bibinfo {author} {\bibfnamefont {D.~W.}\ \bibnamefont
  {Berry}}, \bibinfo {author} {\bibfnamefont {A.~M.}\ \bibnamefont {Childs}},
  \bibinfo {author} {\bibfnamefont {R.}~\bibnamefont {Cleve}}, \bibinfo
  {author} {\bibfnamefont {R.}~\bibnamefont {Kothari}}, \ and\ \bibinfo
  {author} {\bibfnamefont {R.~D.}\ \bibnamefont {Somma}},\ }\bibfield  {title}
  {\bibinfo {title} {Exponential improvement in precision for simulating sparse
  hamiltonians},\ }in\ \href@noop {} {\emph {\bibinfo {booktitle} {Proceedings
  of the forty-sixth annual ACM symposium on Theory of computing}}}\ (\bibinfo
  {year} {2014})\ pp.\ \bibinfo {pages} {283--292}\BibitemShut {NoStop}%
\bibitem [{\citenamefont {Kieferov{\'a}}\ \emph {et~al.}(2019)\citenamefont
  {Kieferov{\'a}}, \citenamefont {Scherer},\ and\ \citenamefont
  {Berry}}]{Kieferov.19}%
  \BibitemOpen
  \bibfield  {author} {\bibinfo {author} {\bibfnamefont {M.}~\bibnamefont
  {Kieferov{\'a}}}, \bibinfo {author} {\bibfnamefont {A.}~\bibnamefont
  {Scherer}}, \ and\ \bibinfo {author} {\bibfnamefont {D.~W.}\ \bibnamefont
  {Berry}},\ }\bibfield  {title} {\bibinfo {title} {Simulating the dynamics of
  time-dependent hamiltonians with a truncated dyson series},\ }\href@noop {}
  {\bibfield  {journal} {\bibinfo  {journal} {Phys. Rev. A}\ }\textbf {\bibinfo
  {volume} {99}},\ \bibinfo {pages} {042314} (\bibinfo {year}
  {2019})}\BibitemShut {NoStop}%
\bibitem [{\citenamefont {Chen}\ \emph {et~al.}(2021)\citenamefont {Chen},
  \citenamefont {Kalev},\ and\ \citenamefont {Hen}}]{Chen.21}%
  \BibitemOpen
  \bibfield  {author} {\bibinfo {author} {\bibfnamefont {Y.-H.}\ \bibnamefont
  {Chen}}, \bibinfo {author} {\bibfnamefont {A.}~\bibnamefont {Kalev}}, \ and\
  \bibinfo {author} {\bibfnamefont {I.}~\bibnamefont {Hen}},\ }\bibfield
  {title} {\bibinfo {title} {Quantum algorithm for time-dependent hamiltonian
  simulation by permutation expansion},\ }\href@noop {} {\bibfield  {journal}
  {\bibinfo  {journal} {PRX Quantum}\ }\textbf {\bibinfo {volume} {2}},\
  \bibinfo {pages} {030342} (\bibinfo {year} {2021})}\BibitemShut {NoStop}%
\bibitem [{\citenamefont {An}\ \emph {et~al.}(2022)\citenamefont {An},
  \citenamefont {Fang},\ and\ \citenamefont {Lin}}]{An.22}%
  \BibitemOpen
  \bibfield  {author} {\bibinfo {author} {\bibfnamefont {D.}~\bibnamefont
  {An}}, \bibinfo {author} {\bibfnamefont {D.}~\bibnamefont {Fang}}, \ and\
  \bibinfo {author} {\bibfnamefont {L.}~\bibnamefont {Lin}},\ }\bibfield
  {title} {\bibinfo {title} {Time-dependent hamiltonian simulation of highly
  oscillatory dynamics and superconvergence for schr{\"o}dinger equation},\
  }\href@noop {} {\bibfield  {journal} {\bibinfo  {journal} {Quantum}\ }\textbf
  {\bibinfo {volume} {6}},\ \bibinfo {pages} {690} (\bibinfo {year}
  {2022})}\BibitemShut {NoStop}%
\bibitem [{\citenamefont {Watkins}\ \emph {et~al.}(2022)\citenamefont
  {Watkins}, \citenamefont {Wiebe}, \citenamefont {Roggero},\ and\
  \citenamefont {Lee}}]{Watkins.22}%
  \BibitemOpen
  \bibfield  {author} {\bibinfo {author} {\bibfnamefont {J.}~\bibnamefont
  {Watkins}}, \bibinfo {author} {\bibfnamefont {N.}~\bibnamefont {Wiebe}},
  \bibinfo {author} {\bibfnamefont {A.}~\bibnamefont {Roggero}}, \ and\
  \bibinfo {author} {\bibfnamefont {D.}~\bibnamefont {Lee}},\ }\bibfield
  {title} {\bibinfo {title} {Time-dependent hamiltonian simulation using
  discrete clock constructions},\ }\href@noop {} {\bibfield  {journal}
  {\bibinfo  {journal} {arXiv:2203.11353}\ } (\bibinfo {year}
  {2022})}\BibitemShut {NoStop}%
\bibitem [{\citenamefont {Mizuta}\ and\ \citenamefont
  {Fujii}(2022)}]{Mizuta.22}%
  \BibitemOpen
  \bibfield  {author} {\bibinfo {author} {\bibfnamefont {K.}~\bibnamefont
  {Mizuta}}\ and\ \bibinfo {author} {\bibfnamefont {K.}~\bibnamefont {Fujii}},\
  }\bibfield  {title} {\bibinfo {title} {Optimal time-periodic hamiltonian
  simulation with floquet-hilbert space},\ }\href@noop {} {\bibfield  {journal}
  {\bibinfo  {journal} {arXiv:2209.05048}\ } (\bibinfo {year}
  {2022})}\BibitemShut {NoStop}%
\bibitem [{\citenamefont {Rajput}\ \emph {et~al.}(2022)\citenamefont {Rajput},
  \citenamefont {Roggero},\ and\ \citenamefont {Wiebe}}]{Rajput.22}%
  \BibitemOpen
  \bibfield  {author} {\bibinfo {author} {\bibfnamefont {A.}~\bibnamefont
  {Rajput}}, \bibinfo {author} {\bibfnamefont {A.}~\bibnamefont {Roggero}}, \
  and\ \bibinfo {author} {\bibfnamefont {N.}~\bibnamefont {Wiebe}},\ }\bibfield
   {title} {\bibinfo {title} {Hybridized methods for quantum simulation in the
  interaction picture},\ }\href@noop {} {\bibfield  {journal} {\bibinfo
  {journal} {Quantum}\ }\textbf {\bibinfo {volume} {6}},\ \bibinfo {pages}
  {780} (\bibinfo {year} {2022})}\BibitemShut {NoStop}%
\bibitem [{\citenamefont {Huang}(2022)}]{Huang.22}%
  \BibitemOpen
  \bibfield  {author} {\bibinfo {author} {\bibfnamefont {H.-Y.}\ \bibnamefont
  {Huang}},\ }\bibfield  {title} {\bibinfo {title} {Learning quantum states
  from their classical shadows},\ }\href@noop {} {\bibfield  {journal}
  {\bibinfo  {journal} {Nature Reviews Physics}\ }\textbf {\bibinfo {volume}
  {4}},\ \bibinfo {pages} {81} (\bibinfo {year} {2022})}\BibitemShut {NoStop}%
\bibitem [{\citenamefont {Verteletskyi}\ \emph {et~al.}(2020)\citenamefont
  {Verteletskyi}, \citenamefont {Yen},\ and\ \citenamefont
  {Izmaylov}}]{Verteletskyi.20}%
  \BibitemOpen
  \bibfield  {author} {\bibinfo {author} {\bibfnamefont {V.}~\bibnamefont
  {Verteletskyi}}, \bibinfo {author} {\bibfnamefont {T.-C.}\ \bibnamefont
  {Yen}}, \ and\ \bibinfo {author} {\bibfnamefont {A.~F.}\ \bibnamefont
  {Izmaylov}},\ }\bibfield  {title} {\bibinfo {title} {Measurement optimization
  in the variational quantum eigensolver using a minimum clique cover},\
  }\href@noop {} {\bibfield  {journal} {\bibinfo  {journal} {The Journal of
  chemical physics}\ }\textbf {\bibinfo {volume} {152}},\ \bibinfo {pages}
  {124114} (\bibinfo {year} {2020})}\BibitemShut {NoStop}%
\bibitem [{\citenamefont {Dyson}(1949)}]{Dyson.49}%
  \BibitemOpen
  \bibfield  {author} {\bibinfo {author} {\bibfnamefont {F.~J.}\ \bibnamefont
  {Dyson}},\ }\bibfield  {title} {\bibinfo {title} {The radiation theories of
  tomonaga, schwinger, and feynman},\ }\href@noop {} {\bibfield  {journal}
  {\bibinfo  {journal} {Physical Review}\ }\textbf {\bibinfo {volume} {75}},\
  \bibinfo {pages} {486} (\bibinfo {year} {1949})}\BibitemShut {NoStop}%
\bibitem [{\citenamefont {Faehrmann}\ \emph {et~al.}(2022)\citenamefont
  {Faehrmann}, \citenamefont {Steudtner}, \citenamefont {Kueng}, \citenamefont
  {Kieferov{\'a}},\ and\ \citenamefont {Eisert}}]{Faehrmann.22}%
  \BibitemOpen
  \bibfield  {author} {\bibinfo {author} {\bibfnamefont {P.~K.}\ \bibnamefont
  {Faehrmann}}, \bibinfo {author} {\bibfnamefont {M.}~\bibnamefont
  {Steudtner}}, \bibinfo {author} {\bibfnamefont {R.}~\bibnamefont {Kueng}},
  \bibinfo {author} {\bibfnamefont {M.}~\bibnamefont {Kieferov{\'a}}}, \ and\
  \bibinfo {author} {\bibfnamefont {J.}~\bibnamefont {Eisert}},\ }\bibfield
  {title} {\bibinfo {title} {Randomizing multi-product formulas for hamiltonian
  simulation},\ }\href@noop {} {\bibfield  {journal} {\bibinfo  {journal}
  {Quantum}\ }\textbf {\bibinfo {volume} {6}},\ \bibinfo {pages} {806}
  (\bibinfo {year} {2022})}\BibitemShut {NoStop}%
\bibitem [{\citenamefont {Yang}\ \emph {et~al.}(2021)\citenamefont {Yang},
  \citenamefont {Lu},\ and\ \citenamefont {Li}}]{Yang.21}%
  \BibitemOpen
  \bibfield  {author} {\bibinfo {author} {\bibfnamefont {Y.}~\bibnamefont
  {Yang}}, \bibinfo {author} {\bibfnamefont {B.-N.}\ \bibnamefont {Lu}}, \ and\
  \bibinfo {author} {\bibfnamefont {Y.}~\bibnamefont {Li}},\ }\bibfield
  {title} {\bibinfo {title} {Accelerated quantum monte carlo with mitigated
  error on noisy quantum computer},\ }\href@noop {} {\bibfield  {journal}
  {\bibinfo  {journal} {PRX Quantum}\ }\textbf {\bibinfo {volume} {2}},\
  \bibinfo {pages} {040361} (\bibinfo {year} {2021})}\BibitemShut {NoStop}%
\bibitem [{\citenamefont {Zeng}\ \emph {et~al.}(2022)\citenamefont {Zeng},
  \citenamefont {Sun}, \citenamefont {Jiang},\ and\ \citenamefont
  {Zhao}}]{Pei.22}%
  \BibitemOpen
  \bibfield  {author} {\bibinfo {author} {\bibfnamefont {P.}~\bibnamefont
  {Zeng}}, \bibinfo {author} {\bibfnamefont {J.}~\bibnamefont {Sun}}, \bibinfo
  {author} {\bibfnamefont {L.}~\bibnamefont {Jiang}}, \ and\ \bibinfo {author}
  {\bibfnamefont {Q.}~\bibnamefont {Zhao}},\ }\bibfield  {title} {\bibinfo
  {title} {Simple and high-precision hamiltonian simulation by compensating
  trotter error with linear combination of unitary operations},\ }\href@noop {}
  {\bibfield  {journal} {\bibinfo  {journal} {arXiv:2212.04566}\ } (\bibinfo
  {year} {2022})}\BibitemShut {NoStop}%
\bibitem [{\citenamefont {Aspuru-Guzik}\ \emph {et~al.}(2005)\citenamefont
  {Aspuru-Guzik}, \citenamefont {Dutoi}, \citenamefont {Love},\ and\
  \citenamefont {Head-Gordon}}]{Aspuru.05}%
  \BibitemOpen
  \bibfield  {author} {\bibinfo {author} {\bibfnamefont {A.}~\bibnamefont
  {Aspuru-Guzik}}, \bibinfo {author} {\bibfnamefont {A.~D.}\ \bibnamefont
  {Dutoi}}, \bibinfo {author} {\bibfnamefont {P.~J.}\ \bibnamefont {Love}}, \
  and\ \bibinfo {author} {\bibfnamefont {M.}~\bibnamefont {Head-Gordon}},\
  }\bibfield  {title} {\bibinfo {title} {Simulated quantum computation of
  molecular energies},\ }\href@noop {} {\bibfield  {journal} {\bibinfo
  {journal} {Science}\ }\textbf {\bibinfo {volume} {309}},\ \bibinfo {pages}
  {1704} (\bibinfo {year} {2005})}\BibitemShut {NoStop}%
\bibitem [{\citenamefont {McArdle}\ \emph {et~al.}(2020)\citenamefont
  {McArdle}, \citenamefont {Endo}, \citenamefont {Aspuru-Guzik}, \citenamefont
  {Benjamin},\ and\ \citenamefont {Yuan}}]{McArdle.20}%
  \BibitemOpen
  \bibfield  {author} {\bibinfo {author} {\bibfnamefont {S.}~\bibnamefont
  {McArdle}}, \bibinfo {author} {\bibfnamefont {S.}~\bibnamefont {Endo}},
  \bibinfo {author} {\bibfnamefont {A.}~\bibnamefont {Aspuru-Guzik}}, \bibinfo
  {author} {\bibfnamefont {S.~C.}\ \bibnamefont {Benjamin}}, \ and\ \bibinfo
  {author} {\bibfnamefont {X.}~\bibnamefont {Yuan}},\ }\bibfield  {title}
  {\bibinfo {title} {Quantum computational chemistry},\ }\href@noop {}
  {\bibfield  {journal} {\bibinfo  {journal} {Reviews of Modern Physics}\
  }\textbf {\bibinfo {volume} {92}},\ \bibinfo {pages} {015003} (\bibinfo
  {year} {2020})}\BibitemShut {NoStop}%
\bibitem [{\citenamefont {Wu}\ \emph {et~al.}(2021)\citenamefont {Wu},
  \citenamefont {Sun}, \citenamefont {Huang},\ and\ \citenamefont
  {Yuan}}]{Wu.21}%
  \BibitemOpen
  \bibfield  {author} {\bibinfo {author} {\bibfnamefont {B.}~\bibnamefont
  {Wu}}, \bibinfo {author} {\bibfnamefont {J.}~\bibnamefont {Sun}}, \bibinfo
  {author} {\bibfnamefont {Q.}~\bibnamefont {Huang}}, \ and\ \bibinfo {author}
  {\bibfnamefont {X.}~\bibnamefont {Yuan}},\ }\bibfield  {title} {\bibinfo
  {title} {Overlapped grouping measurement: A unified framework for measuring
  quantum states},\ }\href@noop {} {\bibfield  {journal} {\bibinfo  {journal}
  {arXiv:2105.13091}\ } (\bibinfo {year} {2021})}\BibitemShut {NoStop}%
\bibitem [{\citenamefont {Magnus}(1954)}]{Magnus.54}%
  \BibitemOpen
  \bibfield  {author} {\bibinfo {author} {\bibfnamefont {W.}~\bibnamefont
  {Magnus}},\ }\bibfield  {title} {\bibinfo {title} {On the exponential
  solution of differential equations for a linear operator},\ }\href@noop {}
  {\bibfield  {journal} {\bibinfo  {journal} {Communications on pure and
  applied mathematics}\ }\textbf {\bibinfo {volume} {7}},\ \bibinfo {pages}
  {649} (\bibinfo {year} {1954})}\BibitemShut {NoStop}%
\bibitem [{\citenamefont {Selinger}()}]{Selinger.12}%
  \BibitemOpen
  \bibfield  {author} {\bibinfo {author} {\bibfnamefont {P.}~\bibnamefont
  {Selinger}},\ }\bibfield  {title} {\bibinfo {title} {Efficient clifford$+t$
  approximation of single-qubit operators},\ }\href@noop {} {\bibinfo
  {journal} {arXiv:1212.6253}\ }\BibitemShut {NoStop}%
\bibitem [{wu_()}]{wu_git}%
  \BibitemOpen
\bibfield  {journal} {  }\href
  {https://github.com/GillianOoO/Overlapped-grouping-measurement} {}\bibinfo
  {note}
  {Https://github.com/GillianOoO/Overlapped-grouping-measurement}\BibitemShut
  {NoStop}%
\end{thebibliography}
%

\newpage

\begin{appendix}
\onecolumngrid

\section{Dyson expansion}\label{app:dy}
The evolution $U=\mathcal{T}\exp\left[-i\int_{\tau_\text{ini}}^{\tau_{\text{ed}}} dtH(t)\right]$ can be expaned with Dyson series as follows 
\begin{align}
U&=\sum_{l=0}^\infty\mathcal{T}\int d\bm{t} (-i)^{l}\prod_{l'=1}^lH(t_{l'}).
\end{align}
Combining with the definition $H(t)=\sum_{p=1}^Ph_p(t)\hat \sigma_p$, we have 
\begin{align}
U&=\sum_{l=0}^\infty\mathcal{T}\int d\bm{t} \sum_{p_1,\cdots,p_l=1}^P\prod_{l'=1}^lh_p(t_{l'})(-i\hat \sigma_p)\\
&=\sum_{l=0}^\infty\sum_{p_1,\cdots,p_l=1}^P\mathcal{T}\int d\bm{t} \prod_{l'=1}^lh_p(t_{l'})\hat P(\bm{p})\\
&=\sum_{l=0}^\infty\sum_{p_1,\cdots,p_l=1}^P\text{Pdy}(l;\bm{p})\hat P(\bm{p})\\
&=e^{\lambda}\sum_{l=0}^{\infty}\text{Poi}(\lambda;l)\left(\sum_{p_1,\cdots,p_l=1}^{P}\text{Pdy}(l;\bm{p})\hat P(\bm{p})\right),\label{eq:DY_app}
\end{align}
which is equivalent to Eq.~\eqref{eq:DY}. The normalization factor for Eq.~\eqref{eq:DY_app} (or Eq.~\eqref{eq:DY}) is
\begin{align}
C_{\text{dy}}=e^{\lambda}\sum_{l=0}^{\infty}\text{Poi}(\lambda;l)\sum_{p_1,\cdots,p_l=1}^{P}|\text{Pdy}(l;\bm{p})|.
\end{align}
We first analysis the summation of $|\text{Pdy}(l;\bm{p})|$ over all $\bm{p}$. Because $h_{p}(t)\geqslant0$ for all $p,t$, we have $\text{Pdy}(l;\bm{p})\geqslant0$. Therefore 
 \begin{align}\label{eq:pdy1}
&\sum_{p_1,p_2,\cdots,p_l=1}^{P}|\text{Pdy}(l;\bm{p})|\notag\\
=&\sum_{p_1,p_2,\cdots,p_l=1}^{P}\text{Pdy}(l;\bm{p})\notag\\
=&\sum_{p_1,p_2,\cdots,p_l=1}^{P}\frac{l!}{\lambda^l}\left(\mathcal{T}\int_{t_{\text{ini}}}^{t_{\text{ed}}}d\bm{t} \prod_{l'=1}^lh_{p_{l'}}(t_{l'})\right)\notag\\
=&\frac{l!}{\lambda^l}\mathcal{T}\int_{\tau_{\text{ini}}}^{\tau_{\text{ed}}}d\bm{t} \prod_{l'=1}^lh_{\text{tot}}(t_{l'}).
\end{align}
Note that we have defined $\mathcal{T}\int_{\tau_{\text{ini}}}^{\tau_{\text{ed}}}=\int_{\tau_{\text{ini}}\leqslant t_l\leqslant t_{l-1}\leqslant  \cdots\leqslant t_1\leqslant\tau_{\text{ed}}}$, and the integration above is invariant under the change of  permutation of $t_{l'}$, i.e. 
\begin{align}\label{eq:pdy2}
\mathcal{T}\int_{\tau_{\text{ini}}}^{\tau_{\text{ed}}}d\bm{t} \prod_{l'=1}^lh_{\text{tot}}(t_{l'}) = \int_{\tau_{\text{ini}}\leqslant t_{\xi(l)}\leqslant t_{\xi(l-1)}\leqslant  \cdots\leqslant t_{\xi(1)}\leqslant\tau_{\text{ed}}}d\bm{t} \prod_{l'=1}^lh_{\text{tot}}(t_{l'})
\end{align}
for all $\xi\in\Xi_l$, where $\Xi_l$ contains all permutations of the indices $\{1,2,\cdots,l\}$. Moreover, the summation of the integrations above over all permutations satisfies
\begin{align}\label{eq:pdy3}
&\sum_{\xi\in\Xi_l}\int_{\tau_{\text{ini}}\leqslant t_{\xi(l)}\leqslant t_{\xi(l-1)}\leqslant  \cdots\leqslant t_{\xi(1)}\leqslant\tau_{\text{ed}}}d\bm{t} \prod_{l'=1}^lh_{\text{tot}}(t_{l'})\notag\\
=&\int_{\tau_{\text{ini}}}^{\tau_{\text{ed}}}\int_{\tau_{\text{ini}}}^{\tau_{\text{ed}}}\cdots\int_{\tau_{\text{ini}}}^{\tau_{\text{ed}}}d\bm{t}\prod_{l'=1}^lh_{\text{tot}}(t_{l'})\notag\\
=&\left(\int_{\tau_{\text{ini}}}^{\tau_{\text{ed}}}dth_{\text{tot}}(t)\right)^l\notag\\
=&\lambda^l.
\end{align}
Combining Eq.~\eqref{eq:pdy2} with Eq.~\eqref{eq:pdy3}, and notice that there are totally $l!$ permutations in $\Xi_l$, we have 
\begin{align}
\left(\mathcal{T}\int_{\tau_{\text{ini}}}^{\tau_{\text{ed}}} d\bm{t} \prod_{l'=1}^lh_{\text{tot}}(t_{l'})\right)\times l!=\lambda^l.\label{eq:pdy4}
\end{align}
Combining Eq.~\eqref{eq:pdy1} with Eq.~\eqref{eq:pdy4}, we have
\begin{align}
\sum_{p_1,p_2,\cdots,p_l=1}^{P}|\text{Pdy}(l;\bm{p})|=\sum_{p_1,p_2,\cdots,p_l=1}^{P}\text{Pdy}(l;\bm{p})=1.\label{eq:prdy5}
\end{align}
So $\text{Pdy}(l;\bm{p})$ is a probability distribution of $\hat P(\bm{p})$. The total normalization factor for LCPS is therefore
\begin{align}
C_{\text{dy}}&= e^\lambda\sum_{l=0}^{\infty}\text{Poi}(\lambda;l)\notag\\
&=e^\lambda\sum_{l=0}^{\infty}\frac{1}{e^\lambda}\frac{\lambda^l}{l!}\notag\\
&=e^\lambda.
\end{align}

\section{Normalization factor for Eq.~\eqref{eq:ULR}}\label{app:ULR}
The normalization factor of remaining order $R$ at Eq.~\eqref{eq:ulrr} is
\begin{align}
C_R&=e^\lambda\sum_{l=2}^\infty\text{Poi}(l;\lambda)\left(\sum_{p_1,\cdots,p_l=1}^{P}\text{Pdy}(l;\bm{p})\right)\notag\\
&=e^\lambda\sum_{l=2}^\infty\frac{1}{e^\lambda} \frac{\lambda^l}{l!}  \notag\\
&=e^{\lambda}-1-\lambda.
\end{align}
 The normalization factor for leading order $L$ in Eq.~\eqref{eq:ulr} is
\begin{align}
C_L=\sum_{p=1}^P\frac{\lambda_p}{\sin\phi}=\sqrt{1+\lambda^2}.
\end{align}
 So the total normalization factor for the leading order rotation is 
 \begin{align}
 C_{\text{lor}}=C_L+C_R&=e^{\lambda}-1-\lambda+\sqrt{1+\lambda^2}\notag\\
 &=1+O(\lambda^2).
 \end{align}
 
\section{Long-time evolution}\label{app:long}
As mentioned in the main text, the long-time evolution is divided into $N_{\text{seg}}$ segments. Each segment $U(\tau_{j},\tau_{j+1})$ is decomposed with leading order rotation technique as Eq.~\eqref{eq:ULR}, and we may express it abstractly as $U(\tau_{j},\tau_{j+1})=C_{\text{lor}}\sum_{s_j}c(s_j)\hat u(s_j)$. The long-time evolution can therefore be expressed as 
\begin{align}
U(0,\tau)&=\prod_{j=N_{\text{seg}}}^1C_{\text{lor}}\sum_{s_j}c(s_j)\hat u(s_j)\notag\\
&=C_{\text{lor}}^{N_{\text{seg}}}\sum_{s_1,s_2,\cdots,s_{N_{\text{seg}}}}\prod_{j=N_{\text{seg}}}^1c(s_j)\hat u(s_j)\notag\\
&=C_{\text{lor}}^{N_{\text{seg}}}\sum_{\bm{s}}c(\bm{s})\hat u(\bm{s})\label{eq:c1}
\end{align}
where we have defined $\bm{s}=[s_1,s_2,\cdots,s_{N_{\text{seg}}}]$, $u(\bm{s})=u(s_{N_\text{seg}})\cdots u(s_{2})u(s_{1})$, and $c(\bm{s})=c(s_{N_\text{seg}})\cdots c(s_{2})c(s_{1})$. Therefore, we can estimate $\langle O\rangle$ with the following quantum circuit,
\begin{center}
\begin{quantikz}[row sep=0.3cm]
\lstick{$|+\rangle$}&\ctrl{1}&\octrl{1}&\meter{$\hat X$}\rstick[wires=2]{$o_m$}\\
\lstick{$|\psi_{\text{ini}}\rangle$}&\gate{\hat u(\bm{s})}\qwbundle{}&\gate{\hat u(\bm{s'})}&\meter{$\hat O$}&
\end{quantikz}
\end{center}
with $\hat u(\bm{s})$ and $\hat u(\bm{s'})$ sampled with probability $c(\bm{s})$ and $c(\bm{s'})$. The circuit above is equivalent to the following circuit with each $u(s_j)$ and $u(s_j)$ sampled independently according to Algorithm.~\ref{alg:2} in the main text.
\begin{center}
\begin{quantikz}[row sep=0.5cm]
\lstick{$|+\rangle$}&\ctrl{1}&\ \ldots\qw&\ctrl{1}&\octrl{1}& \ \ldots\qw &\octrl{1}&\meter{$\hat X$}\rstick[wires=2]{$o_m$}\\
\lstick{$|\psi_{\text{ini}}\rangle$}&\gate{\hat u(s_1)}\qwbundle{}&\ \ldots\qw&\gate{\hat u(s_{N_{\text{seg}}})}&\gate{\hat u(s'_{1})}& \ \ldots\qw&\gate{\hat u(s'_{N_{\text{seg}}})}&\meter{$\hat O$}\\
\end{quantikz}
\end{center}
The estimator of $\langle O\rangle$ is just 
\begin{align}
O_{\text{est}}=C_{\text{long}}^{2}\frac{1}{M}\sum_{m=1}^Mo_m,\label{eq:long_oest}
\end{align}
where $C_{\text{long}}=C_{\text{lor}}^{N_{\text{seg}}}$.
We then discuss the sampling error about Eq.~\eqref{eq:long_oest}. We fix the time integration of total Hamiltonian strength for different segments, i.e. $\int_{\tau_j}^{\tau_{j+1}}h_{\text{tot}}(t)dt=\lambda=$constant. So we have 
\begin{align}
C_{\text{long}}&=\left(\sqrt{1+\lambda^2}+e^\lambda-1-\lambda\right)^{N_{\text{seg}}}.
\end{align}
Let $\Lambda=\int_0^\tau h_{\text{tot}}(t) dt$, we have $\Lambda=N_{\text{seg}}\lambda$, and 
\begin{align}
C_{\text{long}}&=\left(1+O(\lambda^2)\right)^{N_{\text{seg}}}\\
&=\left(1+O(\Lambda^2/N^2_{\text{seg}})\right)^{N_{\text{seg}}}.
\end{align}

According to Eq.~\eqref{eq:err}, with a fixed failure probability, the sampling error for $O_{\text{est}}$ satisfies $|\langle O\rangle-O_{\text{est}}|\leqslant\varepsilon_{\text{samp}}$, where  
\begin{align}
\varepsilon_{\text{samp}}=O\left(\frac{C_{\text{long}}^2}{\sqrt{M}}\right)
=
O\left(\frac{1}{\sqrt{M}}
\left(1+O(\Lambda^2/N^2_{\text{seg}})\right)^{2N_{\text{seg}}}\right).
\end{align}
By setting $N_{\text{seg}}=O(\Lambda^2)$, we have 
\begin{align}
O\left(
\left(1+O(\Lambda^2/N^2_{\text{seg}})\right)^{2N_{\text{seg}}}\right)=\left(1+O(\Lambda^{-2})\right)^{O(\Lambda^2)}=O(1).
\end{align}
Therefore, it suffices to set $M=O(1/\varepsilon_{\text{samp}}^2)$ and $N_{\text{seg}}=O(\Lambda^2)$ to achieve a sampling error smaller than $\varepsilon_{\text{samp}}$.

\section{continuous-qDRIFT~\cite{Berry.20}}\label{app:qd}
Here, we briefly review the c-qDRIFT algorithms, and how the evolution is implemented. Suppose the Hamiltonian is written as 
\begin{align}
H=\sum_{p=1}^{P^{\text{qd}}}H_p(t),
\end{align}
where $e^{-ixH_p(t)}$ can be implemented in the quantum circuit for any $x>0$,
and we should simulate the short-time evolution from $\tau_{\text{ini}}$ to $\tau_{\text{ed}}$. We define a probability distribution (Eq.(77) in~\cite{Berry.20})
\begin{align}
f_p(t)=\frac{\|H_p\|}{\sum_{p'=1}^{P^{\text{qd}}}\int_{\tau_{\text{ini}}}^{\tau_{\text{ed}}}\|H_{p'}\|dt}.\label{eq:fp}
\end{align}
With initial density matrix $\rho_{\text{ini}}$, the output state is given by the following channel 
\begin{align}\label{eq:rqd}
&\mathcal{E}(\tau_{\text{ini}},\tau_{\text{ed}})(\rho_{\text{ini}})\notag\\
=&\sum_{p=1}^{P^{\text{qd}}}\int_{\tau_{\text{ini}}}^{\tau_{\text{ed}}} f_p(t)e^{-iH_p(t)/f_p(t)}\rho_{\text{ini}} e^{iH_p(t)/f_p(t)}dt.
\end{align}
We consider a Pauli decomposition, i.e. $H_{p}(t)= h^{\text{qd}}_p(t)\hat\sigma^{\text{qd}}_p$ with $ h^{\text{qd}}_p(t)\in\mathbb{R}$ and $\hat\sigma^{\text{qd}}_p\in\mathbb{P}$. One of the differences of c-qDRIFT from our URCC method is that the coefficient $h^{\text{qd}}_p(t)$ can be negative. So in general, we have $P/2\leqslant P^{\text{qd}}\leqslant P$, where $P$ is the number of terms in Eq.~\eqref{eq:ham}. 
We have $\|H_p(t)\|=\|h_{p}^{\text{qd}}(t)\hat\sigma_p^{\text{qd}}\|=|h_p^{\text{qd}}(t)|$. Because $\sum_{p=1}^{P^{\text{qd}}}\int_{\tau_{\text{ini}}}^{\tau_{\text{ed}}}|h_p^{\text{qd}}(t)|dt=\lambda$, Eq.~\eqref{eq:fp} can be simplified as 
\begin{align}
f_p(t)=|h_p^\text{qd}(t)|/\lambda,
\end{align}
and Eq.~\eqref{eq:rqd} can be rewritten as 
\begin{align}\label{eq:rqd}
&\mathcal{E}_{\text{qd}}(\tau_{\text{ini}},\tau_{\text{ed}})(\rho_{\text{ini}})\notag\\
=&\sum_{p=1}^P\int_{\tau_{\text{ini}}}^{\tau_{\text{ed}}} \frac{|h_p^{\text{qd}}(t)|}{\lambda} e^{-i\lambda\text{sgn}(h_p^{\text{qd}}(t))\hat \sigma_p}\rho_{\text{ini}} e^{i\lambda\text{sgn}(h_p^{\text{qd}}(t))\hat \sigma_p}dt,
\end{align}
where $\text{sgn}(h_p^{\text{qd}}(t))$ is the sign of $h_p^{\text{qd}}(t)$. Eq.~\eqref{eq:rqd} can be realized by sampling  $t\in[t_{\text{ini}},t_{\text{ed}}]$ and $p$ with probability proportional to $h_p^{\text{qd}}(t)$, and  then apply the unitary $e^{-i\lambda\text{sgn}(h_p^{\text{qd}}(t))\hat \sigma_p}$.

For long-time evolution with $N_{\text{seg}}$ segments, $\mathcal{E}_{\text{qdrift}}(\tau_{j-1},\tau_{j})$ is applied iteratively from $j=1$ to $j=N_{\text{seg}}$. If the initial state is $\rho_{0}$, the final output state of c-qDRIFT method is 
\begin{align}\label{eq:rho_qd}
\rho_\tau^{\text{qd}}=\mathcal{E}_{\text{qd}}(\tau_{N_{\text{seg}}},\tau_{N_\text{seg}-1})\circ\cdots\circ\mathcal{E}_{\text{qd}}(\tau_1,\tau_2)\circ\mathcal{E}_{\text{qd}}(\tau_0,\tau_1)(\rho_{\text{in}}).
\end{align}
It is proved that the algorithmic error of the resulting output state satisfies~\cite{Berry.20}
\begin{align}
\left\|\rho_\tau^{\text{qd}}-\rho_\tau\right\|=O\left(\tilde\Lambda^2/N_{\text{seg}}\right),
\end{align}
where $\rho_{\tau}$ is the ideal quantum state at time $\tau$, and $\tilde\Lambda=\sum_{p=1}^{P^{\text{qd}}}\int_0^\tau\|H_p(t)\|dt$. Because $\|H_p(t)\|=|h_p^{\text{qd}}(t)|$, we have $\tilde\Lambda=\Lambda$. In order words, to achieve a constant algorithmic error level $\varepsilon_{\text{alg}}$, we should set 
\begin{align}
N_{\text{seg}}=O\left(\Lambda^2/\varepsilon_{\text{alg}}\right).
\end{align}
Therefore, the gate count of c-qDRIFT method is $O\left(\Lambda^2n/\varepsilon_{\text{alg}}\right)$ for general Hamiltonian, or $O\left(\Lambda^2k/\varepsilon_{\text{alg}}\right)$ for $k$-local Hamiltonian.

\section{Pauli rotations}\label{app:r}
The exponential of Pauli string is an elementary operation in our algorithm.
Without loss of generality, we consider an operator 
\begin{align}\label{eq:r}
\hat r=\exp(-i\phi \hat{\widetilde\sigma}_1\otimes\hat{\widetilde\sigma}_2\otimes\cdots\otimes\hat{\widetilde\sigma}_n)
\end{align}
where, $\widetilde\sigma_j\in\{\hat I,\hat X,\hat Y,\hat Z\}$ is the single-qubit Pauli operator applied at the $j$th qubit, and $\phi\in\mathbb{R}$. Suppose there are totally $k$ non-identity Pauli operators among $\hat{ \widetilde\sigma_j}$, and we denote $j'_s$ as the index of the $s$th non-identity Pauli operators. For example, if $\sigma=\hat I\otimes \hat X\otimes \hat I\otimes \hat Y\otimes \hat I\otimes \hat Z$, we have $k=3$, and $j'_1=2$, $j'_2=4$, $j'_3=6$. We denote CNOT$(j_\text{c},j_\text{t})$ as the two-qubit controlled-not gate with the $j_{\text{c}}$th qubit as control qubit, and $j_{\text{t}}$th qubit as target qubit. We define 
\begin{align}
  W(\hat X)=\frac{1}{\sqrt{2}}\begin{pmatrix}1&1\\1&-1\end{pmatrix},&\quad W(\hat Y)=\begin{pmatrix}1&-i\\-i&1\end{pmatrix},
  \quad W(\hat Z)=\begin{pmatrix}1&0\\0&1\end{pmatrix}, \quad \text{Ph}(\theta)=\begin{pmatrix}e^{-i\theta}&0\\0&e^{i\theta}\end{pmatrix}.
\end{align}
The Eq.~\eqref{eq:r} can be decomposed into single- and two-qubit gates as follows
\begin{align}
\hat r=&\left(\prod_{s=1}^{k}W(\sigma_s)^\dag\right)\left(\prod_{s=1}^{k-1}\text{CNOT}(j'_{s},j'_{s+1})\right)\text{Ph}(\theta)\left(\prod_{s=1}^{k-1}\text{CNOT}(j'_{k-s},j'_{k-s+1})\right)\left(\prod_{s=1}^{k}W(\sigma_s)\right).
\end{align}
For example, if $k=3$, we have:
\begin{center}
\begin{quantikz}
&\gate{\hat r}\qwbundle{} & \qw
\end{quantikz}
$=$
\begin{quantikz}
\lstick{$j_1$}&\gate{W(\hat{\tilde{\sigma}}_1)}&\ctrl{1}&\qw&\qw&\qw&\ctrl{1}&\gate{W(\hat{\tilde{\sigma}}_1)^\dag}&\qw\\
\lstick{$j_2$}&\gate{W(\hat{\tilde{\sigma}}_2)}&\targ{}&\ctrl{1}&\qw&\ctrl{1}&\targ{}&\gate{W(\hat{\tilde{\sigma}}_2)^\dag}&\qw\\
\lstick{$j_3$}&\gate{W(\hat{\tilde{\sigma}}_3)}&\qw&\targ{}&\gate{\text{Ph}(\phi)}&\targ{}&\qw&\gate{W(\hat{\tilde{\sigma}}_3)^\dag}&\qw
\end{quantikz}
\end{center}

For rotation $\hat r$ controlled by state $|1\rangle_{\text{anc}}\langle1|$ of an ancillary qubit, the quantum circuit is as follows.
\begin{center}
\begin{quantikz}
\lstick{data qubits} &\gate{\hat r}\qwbundle{} & \qw\\
\lstick{anc}&\ctrl{-1}&\qw&
\end{quantikz}
$=$
\begin{quantikz}[column sep=.2cm]
\lstick{$j_1$}&\gate{W(\hat{\tilde{\sigma}}_1)}&\ctrl{1}&\qw&\qw&\qw&\qw&\qw&\qw&\ctrl{1}&\gate{W(\hat{\tilde{\sigma}}_1)^\dag}&\qw\\
\lstick{$j_2$}&\gate{W(\hat{\tilde{\sigma}}_2)}&\targ{}&\ctrl{1}&\qw&\qw&\qw&\qw&\ctrl{1}&\targ{}&\gate{W(\hat{\tilde{\sigma}}_2)^\dag}&\qw\\
\lstick{$j_3$}&\gate{W(\hat{\tilde{\sigma}}_3)}&\qw&\targ{}&\gate{\text{Ph}(\frac{\phi}{2})}&\targ{}&\gate{\text{Ph}(\frac{\phi}{2})^\dag}&\targ{}&\targ{}&\qw&\gate{W(\hat{\tilde{\sigma}}_3)^\dag}&\qw\\
\lstick{anc}&\qw&\qw&\qw&\qw&\ctrl{-1}&\qw&\ctrl{-1}&\qw&\qw&\qw&\qw
\end{quantikz}
\end{center}

For rotation $\hat r$ controlled by $|0\rangle_{\text{anc}}\langle0|$ of the ancillary qubit, the circuit is similar.
\section{Clifford+$T$ decomposition}\label{app:ct}
We have shown that with ideal single- and two-qubit gates, the gate count of our protocol is independent of the accuracy. In the error-corrected quantum computing setting, however, we should decompose the circuit into elementary gates that are fault-tolerant. For example, surface-code-based quantum computation requires the decomposition of the circuit into Clifford+$T$ gates. The Clifford+$T$ decompositions are in general non-ideal, although the error rate reduces rapidly with gate count. Below, we analyze the Clifford+$T$ gate count for our protocol.

In our quantum circuit, except for the phase gate Ph$(\phi/2)$, all other elementary single- and two-qubit gates can be decomposed into constant number of Clifford+$T$ gates without error. According to App.~\ref{app:r}, there are totally $O(\Lambda^2)$ number of phase gates in our protocol. Suppose each phase gate are realized with $O(N_{\text{ph}})$ Clifford+$T$ gates, the total gate count is   $O(\Lambda^2(k+N_{\text{ph}}))$. $N_{\text{ph}}$ depends on the error rate we want to achieve. According to~\cite{Selinger.12}, suppose the error of each phase gate is $\varepsilon_{\text{ph}}$, there exist a decomposition protocol satisfying $N_{\text{ph}}=O(\log(1/\varepsilon_{\text{ph}}))$. Because there are totally $O(\Lambda^2)$ phase gates, the total algorithmic error is $\varepsilon_{\text{alg}}=O(\Lambda^2\varepsilon_{\text{ph}})$. To achieve this error rate, it suffices to set $N_{\text{ph}}=O(\log(\Lambda/\varepsilon_{\text{ph}}))$. Therefore, the total Clifford+$T$ gate count is $O(\Lambda^2(k+\log(\Lambda/\varepsilon_{\text{alg}})))$.

In Fig.~\ref{fig:num}(e) and (f) for $T$ count ratio, we have assumed that the $\varepsilon_{\text{ph}}$ is sufficiently small, such that the $T$ count for elementary gates other than phase gates can be neglected. If $N_{\text{seg}}=N_{\text{seg},\text{urcc}}$ for URCC method and $N_{\text{seg}}=N_{\text{seg},\text{qd}}$ for c-qDRIFT method, there are totally $4N_{\text{seg},\text{urcc}}$ and $N_{\text{seg},\text{qd}}$ phase gates respectively (App.~\ref{app:r}). The $T$ count ratio is therefore estimated as $N_{\text{seg},\text{qd}}/4N_{\text{seg},\text{urcc}}$ for Fig.~\ref{fig:num}(e) and (f).

\section{Combination URCC with grouping measurement}\label{app:group}
When a group of observables is compatible with each other, they can be measured simultaneously. For example, $\hat Z\otimes \hat I$ and $\hat I\otimes \hat Z$ can be measured simultaneously with the observable $\hat Z\otimes \hat Z$. Quantum circuit in Fig.~\ref{fig:1}(b) is compatible with the simultaneous measurement techniques~\cite{Pei.22}. Here, we discuss the sampling error of non-overlapped grouping in our URCC protocol~\cite{Verteletskyi.20}. The generalization to overlapped grouping~\cite{Wu.21} is straightforward.

Suppose $\hat Q=\bigotimes_{i=1}^n\hat Q_i$ and $\hat R=\bigotimes_{i=1}^n\hat R_i$ with $\hat Q_i,\hat R_i\in\{\hat I,\hat X,\hat Y,\hat Z\}$, we denote $\hat Q \triangleright\hat R$ when  $\hat Q_i=\hat R_i$ or  $\hat  Q_i=\hat I$ for all $i$. For qubit systems, the target observable can generally be decomposed as LCPS
\begin{align}
\hat O=\sum_{k=1}^K\alpha_k\hat O^{(k)},
\end{align}
where $\alpha_k\in\mathbb{R}$, $\hat O^{(k)}\in\mathbb{P}$. All components $\mathcal{O}=\{O^{(k)}\}$ are partitioned into $G$ groups $e_1,e_2,\cdots,e_G$, satisfying $\bigcup_{g=1}^Ge_g=\mathcal{O}$ and $e_g\cap e_{g'}=\varnothing$ for $g\neq g'$. In each group, there exist an observable $\hat R_g$ such that  $\hat Q\triangleright \hat R_g$ for $\forall \hat Q\in e_j$. The way of grouping is not unique, and finding the optimal grouping strategy is NP-hard. But there are heuristic grouping strategies with polynomial runtime, such as the largest degree first (LDF) grouping method~\cite{Verteletskyi.20}.

Similar to single observable measurement, we suppose the evolution is decomposed as Eq.~\eqref{eq:normf} in the main text. For each $R_g$, we implement the quantum circuit below for $M_g>0$ times, 
\begin{center}
\begin{quantikz}[row sep=0.3cm]
\lstick{$|+\rangle$}&\ctrl{1}&\octrl{1}&\meter{$\hat X$}\\
\lstick{$|\psi_{\text{ini}}\rangle$}&\gate{\hat u(s)}\qwbundle{}&\gate{\hat u(s')}&\meter{$\hat R_g$}&
\end{quantikz}
\end{center}
where $u(s)$ and $u(s')$ are sampled according to the probability $c(s)$ and $c(s')$.
We denote the measurement outcome at the $m$th measurement of $R_g$ as $\bm{r_{g,m}}\in\{-1,+1\}^{n+1}$, including ancillary and data qubits.

 Let $M=\sum_{g=1}^GM_g$, and
\begin{align}
o_{g,m}=\sum_{\hat O^{(k)}\in e_g}\alpha_k\frac{M}{M_g}\mu(\bm{r_{g,m}},\text{supp}(\hat X\otimes\hat O^{(k)})).
\end{align}
Here, $\text{supp}(\hat X\otimes\hat O^{(k)})$ contains all qubits that $\hat O^{(k)}$ acting nontrivially on it. For example, $\text{supp}(\hat X\otimes\hat Y\otimes\hat I\otimes\hat Z)=\{1,2,4\}$. Moreover, $\mu(\bm{r_{g,m}},\text{supp}(\hat O^{(k)}))=\prod_{j\in\text{supp}(\hat O^{(k)})}r_{g,m,j}$, where $r_{g,m,j}$ is the measurement outcome of the $j$th qubit for $r_{g,m}$. 

The estimator of $\langle O\rangle$ in the grouping measurement scenario defined as 
\begin{align}
O_{\text{est}}=\frac{C^2}{M}\sum_{g=1}^G\sum_{m=1}^{M_g}o_{g,m},
\end{align}
which satisfies $\overline {O_{\text{est}}}=\langle O\rangle$. To minimize the variance of $O_{\text{est}}$,  $M_g$ is set to be proportional to the $l_1$-norm of $e_g$~\cite{Verteletskyi.20,Wu.21}.

 Similar to single observable measurement, we also estimate the sampling error with Hoeffding's bound.
We define the range of the measurement outcome for $R_g$ as $\|R_g\|_r\equiv\max(o_{g,m})-\min(o_{g,m})$, where $\max(o_{g,m})$ and $\min(o_{g,m})$ represents the largest and smallest possible values of $o_{g,m}$. Note that $\max(o_{g,m})$ and $\min(o_{g,m})$ corresponds to the largest and smallest eigenvalues of $\sum_{\hat Q^{(k)}\in e_g}\alpha_k\frac{M}{M_g}\hat Q^{(k)}$.

Because $o_{g,m}$ are independent of each other, according to Hoeffding's bound, we have
\begin{align}
&\text{Pr}\left[|O_{\text{est}}-\langle \tilde O\rangle|\geqslant \varepsilon_{\text{samp}}\right]\leqslant2\exp\left(\frac{-2M^2\varepsilon_{\text{samp}}^2}{C^4\sum_{g=1}^{G}M_g\|R_g\|_r^2}\right)
\end{align}
Therefore, with probability $\delta\in(0,1)$, the sampling error satisfies $|O_{\text{est}}-\langle O\rangle|\leqslant\varepsilon_{\text{samp}}$ where
\begin{align}
 \varepsilon_{\text{samp}}=\frac{C^2}{M}\sqrt{\frac{1}{2}\ln(2/\delta)\sum_{g=1}^GM_g\|\hat R_g\|_r^2}.
 \end{align}

\section{Details about numerical simulations}\label{app:num}
\subsection{Errors for c-qDRIFT}
To be consistent with the the URCC method, we estimate the sampling error for c-qDRIFT also with Hoeffding's bound. The mean value of sampling is $\overline{O_{\text{est}}}=\text{Tr}(\rho_{\tau}^{\text{qd}}\hat O)$, where $\rho_{\tau}^{\text{qd}}$ is defined in Eq.~\eqref{eq:rho_qd}. If we want the total error, containing both algorithmic error and sampling error, to be within $\varepsilon$, the failure probability, satisfies
\begin{align}
\delta&=\text{Pr}\left[\left|O_{\text{est}}-\langle O\rangle\right|\geqslant \varepsilon\right]\\
&=\text{Pr}\left[O_{\text{est}}-\langle O\rangle\geqslant \varepsilon\right]+\text{Pr}\left[O_{\text{est}}-\langle O\rangle\leqslant -\varepsilon\right]\\
&\leqslant \exp\left(\frac{-M(\varepsilon-\varepsilon_{\text{alg}})^2}{2\|\hat O\|^2}\right)+\exp\left(\frac{-M(\varepsilon+\varepsilon_{\text{alg}})^2}{2\|\hat O\|^2}\right)
\end{align}
where $\varepsilon_{\text{alg}}=\overline{O_{\text{est}}}-\langle O\rangle$ is the algorithmic error. In our simulation, we first calculate $\varepsilon_{\text{alg}}$, and the value of total error $\varepsilon$ is chosen such that $\delta=\exp\left(\frac{-M(\varepsilon-\varepsilon_{\text{alg}})^2}{2\|\hat O\|^2}\right)+\exp\left(\frac{-M(\varepsilon+\varepsilon_{\text{alg}})^2}{2\|\hat O\|^2}\right)$ is satisfied with $\delta=0.05$. For group measurement, the process is similar.

\subsection{Details about adiabatic preparation of the ground state of $H_2$}
In obtaining the Hamiltonian of $H_2$, Jordan-Wigner transformation with the basis set STO-3G is used. The exact form of the Hamiltonian and the code for generating measurement protocol is referred to~\cite{wu_git}.

\end{appendix}

\end{document}